\documentclass[a4paper,11pt]{article}
\usepackage[utf8]{inputenc}
\usepackage[english]{babel}

\usepackage{graphicx}
\graphicspath{ {img/}, {plots/} }
\usepackage{booktabs}
\usepackage{tabularx}
\usepackage{multirow}
\usepackage{svg}

\usepackage{soul}
\usepackage[dvipsnames]{xcolor}

\usepackage{pgfplots}
\pgfplotsset{compat=1.18}

\usepackage{comment}

\usepackage{array} 
\usepackage{anyfontsize}

\usepackage{makecell}

\usepackage{fancyhdr}
\usepackage{datetime} 

\usepackage{subcaption}
\usepackage{needspace} 
\usepackage{pdflscape}
\usepackage{authblk}
\usepackage{setspace}

\usepackage[a4paper,margin=2cm]{geometry}

\usepackage{tikz}
\usetikzlibrary{positioning,
                calc, 
                chains}

\usepackage[
    backend=biber,
    style=numeric-comp,
    sorting=none
]{biblatex}
\AtEveryCitekey{\clearfield{issue}}

\usepackage[colorlinks=true, allcolors=blue]{hyperref}

\newcommand{\newparagraph}[1]{
    \needspace{3\baselineskip}
    \vspace{\baselineskip}
    \noindent\textbf{#1}
    \par\noindent\ignorespaces
}

\title{Analyzing Vaccine Manufacturing Supply Chain Disruptions for Pandemic Preparedness using Discrete-Event Simulation}

\author[1]{Robin Kelchtermans}
\author[1]{Valentijn Stienen}
\author[1,3]{Guido Dietrich}
\author[2,3]{Mauro Bernuzzi}
\author[1]{Nico Vandaele}

\affil[1]{Access-To-Medicines Research Centre, Faculty of Economics and Business, KU Leuven, Belgium}
\affil[2]{Università Cattolica del Sacro Cuore, Milan, Italy}
\affil[3]{Coalition for Epidemic Preparedness Innovations (CEPI)}

\date{}

\begin{document}

\onehalfspacing

\maketitle

\begin{abstract}
The COVID-19 pandemic exposed critical vulnerabilities in vaccine supply chains, highlighting the need for robust manufacturing for rapid pandemic response  to support CEPI’s 100 Days Mission. We develop a discrete-event simulation model to analyze supply chain disruptions and enables policymakers and vaccine manufacturers to quantify disruptions and assess mitigation strategies. Unlike prior studies examining components in isolation, our approach integrates production processes, quality assurance and control (QA/QC) activities, and raw material procurement to capture system-wide dynamics. A detailed mRNA case study analyzes disruption scenarios for a facility targeting 50 million doses: facility shutdowns, workforce reductions, raw material shortages, infrastructure failures, extended procurement lead times, and increased QA/QC capacity. Three main insights emerge. First, QA/QC personnel are the primary bottleneck, with utilization reaching 84.5\% under normal conditions while machine utilization remains below 33\%. Doubling QA/QC capacity increases annual output by 79.1\%, offering greater returns than equipment investments. Second, raw material disruptions are highly detrimental, with extended lead times reducing three-year output by 19.6\% and causing stockouts during 51.8\% of production time. Third, the model shows differential resilience: acute disruptions ( workforce shortages, shutdowns, power outages) allow recovery within 6 to 9 weeks, whereas chronic disruptions (supply delays) cause prolonged performance degradation.
\end{abstract}

\noindent\textbf{Keywords:} vaccine manufacturing, discrete-event simulation, pandemic preparedness, mRNA technology platform, supply chain resilience\\

\noindent \textbf{Corresponding author:} Robin Kelchtermans, robin.kelchtermans@kuleuven.be

\section{Introduction}
\label{s:introduction}
Vaccines have been instrumental in advancing global public health by preventing the spread of infectious diseases, saving an estimated 154 million lives in the last 50 years \cite{Doherty2016,WorldHealthOrganization2024}. While most countries have established immunization programs, the COVID-19 pandemic exposed critical vulnerabilities in vaccine manufacturing and supply chain systems \cite{Plotkin2017, Heilweil2021, Wouters2021}. The pandemic revealed that missing even one critical raw material, such as bioreactor bags or filters, could delay entire production processes, leading to significant delays in vaccine availability and distribution \cite{Rodgers2021}. These disruptions highlighted fundamental challenges including manufacturing capacity constraints, raw material procurement difficulties from specialized suppliers with long lead times, quality assurance bottlenecks due to limited skilled personnel for testing and regulatory approval processes, and global distribution inequities \cite{Plotkin2017, Heilweil2021, Wouters2021, Rodgers2021}.

In response to these challenges, international initiatives have established ambitious pandemic preparedness targets. The Coalition for Epidemic Preparedness Innovations (CEPI) launched the ``100 Days Mission", aiming to enable the development and deployment of effective vaccines within 100 days of pandemic threat declaration \cite{Saville2022}. 
Similarly, the European Commission's Health Emergency Preparedness and Response Authority (HERA) has set a target to ensure capacity for producing 325 million vaccine doses annually during public health emergencies, distributed across six manufacturers and three different technologies (viral vector, recombinant protein and mRNA), with varying capacities for each manufacturer \cite{ECHaDEA2023}. 
Additionally, Germany's Center for Pandemic Vaccines and Therapeutics (ZEPAI) focuses on strengthening pandemic preparedness through coordinated research, development of platform technologies, and establishment of rapid response capabilities for future health emergencies \cite{Paul-Ehrlich-Institut2025}.
Achieving these objectives requires robust and resilient vaccine manufacturing supply chains, with particular attention to timely procurement of critical raw materials and availability of skilled quality assurance and quality control (QA/QC) personnel to ensure vaccine safety and regulatory compliance.

Vaccine manufacturing supply chains present unique challenges that differentiate them from traditional industrial supply chains and require specialized analysis approaches. Unlike conventional manufacturing, vaccine manufacturing is highly regulated, involves complex biological processes with inherent variability, and operates under stringent quality requirements that cannot be compromised even during emergencies \cite{Plotkin2017, Duijzer2018}. These supply chains are characterized by horizontal complexity, requiring coordination across numerous specialized raw material suppliers rather than vertical supply chain tiers \cite{Bode2015}. Critical raw materials (e.g., specialized lipids, and single-use bioreactor bags) are often sourced from a limited number of qualified suppliers with long and uncertain lead times \cite{Feddema2023}. Additionally, QA/QC activities require highly skilled personnel who cannot be rapidly scaled during emergencies, creating persistent bottlenecks that extend batch lead times regardless of production capacity \cite{Plotkin2017, Feddema2023}.

We want to examine the impact of disruptions in vaccine manufacturing systems and their cascading effects on pandemic response capabilities. Understanding how disruptions propagate through vaccine manufacturing systems is critical for developing effective mitigation strategies and ensuring robust pandemic preparedness. Conventional analytical approaches prove to be insufficient for this analysis due to the complex interactions between production processes, QA/QC resource constraints, and raw material availability that determine system-wide performance during disruptions. 
Traditional inventory models and multi-echelon inventory optimization techniques, such as those developed by \textcite{Graves2000}, assume stationary demand patterns and deterministic supply conditions that are fundamentally incompatible with pandemic scenarios characterized by sudden demand surges and frequent supply disruptions \cite{DeKok2018}. Similarly, closed-form queueing models cannot represent the dynamic interactions between production processes, QA/QC resource constraints, and raw material availability that determine system-wide performance during disruptions.
The stochastic nature of raw material lead times, resource constraints in QA/QC activities, and feedback loops between production stages create complex system behaviors where small disruptions can have amplified effects, requiring sophisticated modeling approaches to capture realistic pandemic response dynamics.

This paper addresses these challenges by developing and applying a discrete-event simulation model to examine how raw material procurement disruptions and QA/QC workforce constraints interact to affect vaccine manufacturing performance during pandemic response scenarios. Through fine-grained discrete-event simulation modeling that integrates production processes, QA/QC activities, and raw material procurement, we aim to: (1) identify critical bottlenecks in vaccine manufacturing under normal and disrupted conditions, (2) quantify the impact of various disruption scenarios on production performance and pandemic response capabilities, and (3) evaluate the effectiveness of potential mitigation strategies for enhancing supply chain resilience. 

%---

\subsection{Literature review}
The complexity of vaccine manufacturing supply chains, particularly their vulnerability to disruptions during pandemic scenarios, has prompted researchers to explore various analytical approaches for understanding and optimizing these systems. This literature review examines existing research across four key areas: traditional supply chain modeling approaches and their limitations for pandemic scenarios, the unique characteristics and complexities of vaccine manufacturing supply chains, supply chain disruption and resilience research including recent pandemic-related studies, and simulation modeling applications in bio-pharmaceutical manufacturing. 

\subsubsection{Traditional supply chain modeling approaches and their limitations}

Classical supply chain optimization models have been extensively developed for managing multi-echelon inventory systems and production planning. \textcite{Graves2000} established foundational approaches for safety stock optimization across supply chain tiers by focusing on cost minimization under known, stationary demand patterns and deterministic lead times \cite{DeKok2018}.

However, these traditional approaches exhibit limitations when applied to vaccine manufacturing during pandemic conditions. First, they rely on stationary demand assumptions that are incompatible with the sudden demand surges characteristic of health emergencies. Second, they prioritize cost optimization over response speed and availability, which are primary concerns during pandemic response. Third, they assume deterministic or well-characterized stochastic processes, while pandemic scenarios involve uncertainty and cascading disruptions that violate standard modeling assumptions \cite{DeKok2018}.

Multi-echelon inventory optimization approaches, while more sophisticated, focus primarily on vertical integration across supply chain tiers rather than the horizontal complexity that characterizes vaccine manufacturing \cite{Bode2015}. Vaccine supply chains require coordination across numerous specialized raw material suppliers with distinct lead times and quality requirements, rather than traditional hierarchical supplier relationships, creating fundamentally different optimization challenges.

\subsubsection{Vaccine manufacturing complexity and supply chain characteristics}

Vaccine manufacturing presents unique supply chain characteristics that differentiate it from traditional manufacturing contexts. \textcite{Bode2015}'s classification framework identifies vaccine supply chains as exhibiting high horizontal complexity due to their dependence on numerous specialized raw material suppliers, each operating under distinct regulatory requirements, quality standards, and capacity constraints that limit substitutability during disruptions \cite{Feddema2023}.

Vaccine manufacturing shares structural similarities with assemble-to-order (ATO) systems, where components are pre-positioned and final assembly occurs upon demand \cite{Song2003}. However, vaccine manufacturing differs significantly from traditional ATO systems due to unpredictable epidemic outbreaks demand surges, extensive regulatory QA/QC requirements, and time-critical public health constraints. These unique characteristics make traditional ATO inventory optimization approaches less applicable to vaccine manufacturing scenarios, as they cannot capture the regulatory complexity and quality constraints that dominate vaccine manufacturing operations.

The regulatory environment creates additional analytical challenges through stringent quality requirements that cannot be relaxed even during health emergencies. \textcite{Plotkin2017} emphasizes that QA/QC processes in vaccine manufacturing involve multiple sequential and parallel testing stages, each requiring specialized personnel with specific qualifications that cannot be rapidly scaled. This regulatory complexity distinguishes vaccine supply chains from traditional manufacturing systems where quality control can often be streamlined or accelerated during emergency production scenarios.

\textcite{Duijzer2018} identifies vaccine manufacturing as particularly vulnerable to supply chain disruptions due to the combination of biological process variability, temperature-controlled storage requirements, and limited shelf life that constrain inventory management strategies. Unlike traditional manufacturing where safety stock can be maintained indefinitely, vaccine manufacturing involves time-sensitive intermediate products and finished goods that require continuous turnover, limiting the effectiveness of traditional inventory buffers against supply disruptions.

\subsubsection{Supply chain disruption and resilience research}

The COVID-19 pandemic revealed specific vulnerabilities in vaccine manufacturing capacity scaling, with \textcite{Feddema2023} and \textcite{Rele2021} identifying key bottlenecks including production facility limitations, resource scarcity, and regulatory alignment challenges. Their findings also demonstrated how disruptions can create interdependencies that amplify impacts beyond individual failure points, requiring analytical approaches capable of capturing system-wide effects during crisis scenarios.

\textcite{Ivanov2020} provides extensive coverage of simulation approaches in supply chain management, emphasizing their applicability to healthcare contexts while identifying limited research specifically addressing pharmaceutical supply chain dynamics during pandemic conditions. Recent work by \textcite{Belin2025} confirms that operations research applications in healthcare rarely address pandemic preparedness specifically, with most studies examining isolated production aspects rather than integrated system-wide interactions.

\textcite{Ivanov2021b} propose digital twins for modeling complex supply chains to enable real-time policy adjustments, representing a promising direction for supply chain resilience analysis. However, existing applications remain limited in scope and do not address the integrated challenges of vaccine manufacturing under disruption scenarios, particularly the regulatory complexity and quality constraints that characterize bio-pharmaceutical production.

\textcite{Sawik2025} develops mathematical optimization frameworks for economically viable reshoring decisions under ripple effect conditions, balancing operational costs with disruption risks. Their work demonstrates that reshoring becomes economically justified when disruption probability multiplied by expected disruption costs exceeds the cost differential between domestic and offshore suppliers, particularly when ripple effects from upstream disruptions compound throughout the supply network.
This economic optimization perspective is particularly relevant for vaccine manufacturing, where critical raw materials sourced from distant specialized suppliers create vulnerabilities to cascading disruptions that can halt production entirely. However, their focus on post-disruption cost optimization and reshoring as a reactive strategy does not address the proactive simulation capabilities needed for pandemic preparedness planning, where the ability to evaluate supply chain configurations and resilience investments before disruptions occur is essential for effective policy design and resource allocation decisions.

The reviewed literature reveals limited research specifically addressing integrated supply chain analysis for pandemic preparedness scenarios, highlighting a gap in prospective modeling capabilities that would enable evaluation of alternative strategies and policies before real disruptions occur.

\subsubsection{Simulation modeling in bio-pharmaceutical manufacturing}

Simulation modeling has been applied to various aspects of pharmaceutical manufacturing, though typically focusing on isolated processes rather than integrated supply chain systems. For example, \textcite{Akmayan2024} demonstrates detailed process simulations for recombinant protein production using commercial tools like SuperPro Designer, showing how simulation can estimate production costs and process parameters. However, such studies operate in isolation from broader supply chain dynamics, missing critical interactions between production, procurement, and quality control and assurance activities. For an extended overview on discrete-event simulation (in bio-pharmaceutical manufacturing), we refer to \textcite{Law2015}.
\\

\noindent
To the best of our knowledge, no existing studies have integrated raw material procurement and QA/QC activities within vaccine manufacturing supply chains. This represents a significant gap, as these factors are significant contributors to vaccine manufacturing and release lead times, and can represent bottlenecks essential for accurately assessing supply chain resilience under disruption scenarios. Our work addresses this gap by developing a simulation model that captures the dynamic interplay between raw material availability, procurement lead times, QA/QC workforce limitations, and the inherent uncertainties in vaccine manufacturing processes, providing a more realistic foundation for evaluating pandemic preparedness strategies.

\subsection{Contributions}
This paper makes three key contributions: 
(1) the development of the first discrete-event simulation model that integrates production processes, QA/QC activities and raw material procurement within bio-pharmaceutical manufacturing, addressing a critical gap in existing studies that examine these components in isolation. Unlike previous analytical approaches that rely on deterministic assumptions, our holistic simulation model captures stochastic disruptions, cascading effects, and dynamic interactions between system components, providing the first quantitative assessment of trade-offs between resilience investments and operational performance in vaccine supply chains during pandemic conditions. 
(2) A comprehensive decision-support system for analyzing different scenarios. This system enables policymakers and manufacturers to quantitatively evaluate disruption impacts and mitigation strategies under realistic operational constraints, providing actionable insights for pandemic preparedness planning and supply chain investment decisions.
And (3) a detailed mRNA vaccine case study that demonstrates the practical applicability of the integrated modeling approach and reveals the critical bottlenecks of QA/QC personnel capacity and raw material procurement in pandemic response scenarios. 

% ========================================================================================

\section{Methodology}
\label{s:methodology}
This section presents the methodology used to assess the impact of disruptions on vaccine manufacturing supply chains. We first explain the selection of discrete-event simulation as the most appropriate modeling approach for this research domain. Subsequently, we describe the simulation model structure, including the production processes, raw materials, and QA/QC activities. Finally, we detail the model implementation, experimental design, and performance metrics used to evaluate different disruption scenarios.

\subsection{Discrete-event simulation}
Discrete-event simulation (DES) is the most suitable methodological approach for analyzing the impact of disruptions on vaccine manufacturing due to several key characteristics that align with the unique challenges of vaccine supply chains.

First, vaccine manufacturing is characterized by discrete events that occur at specific points in time, such as batch completions, raw material arrivals, or QA/QC approvals. These events trigger state changes and affect subsequent activities, making DES the natural choice for capturing this event-driven behavior. \textcite{Law2015} identifies discrete-event simulation as particularly suitable for pharmaceutical manufacturing due to its ability to capture batch-oriented processes, resource contention, and stochastic variability in complex manufacturing environments.

Second, the stochastic nature of vaccine manufacturing processes, including variable processing times, uncertain raw material lead times, and varying yield outputs, requires a modeling approach that can explicitly represent probabilistic distributions and their interactions. 

Third, resource contention is fundamental to vaccine manufacturing, where limited QA/QC personnel, production machines, and storage capacity must be allocated among competing demands. DES naturally handles these constraints through queuing mechanisms and resource allocation algorithms, enabling the model to capture bottlenecks and their dynamic evolution over time.

Furthermore, DES provides the flexibility needed to model diverse disruption scenarios relevant to pandemic preparedness, from short-term acute events to chronic systemic issues. The modular nature of DES allows for easy modification of parameters and structure to represent different disruption types, making it an ideal platform for scenario analysis and policy evaluation.

Alternative modeling approaches have significant limitations for this application. Analytical models, while computationally efficient, cannot capture the complex interactions and non-linear behaviors that characterize disrupted vaccine supply chains. Agent-based models, though capable of representing individual decision-making, are not necessary for this research as the focus is on system-level performance rather than emergent behaviors from autonomous agents. System dynamics models, while suitable for strategic planning, lack the operational detail needed to evaluate specific disruption scenarios and mitigation strategies.

Our model is implemented as a digital model \cite{Kritzinger2018}, providing a virtual representation of the vaccine manufacturing system that can be used for experimentation and analysis without the costs and risks associated with real-world testing. This simulation model can also be used as a pro-active decision-support system, enabling stakeholders to evaluate potential disruptions and mitigation strategies before they occur in practice. As discussed by \textcite{Broekaert2025}, having a model ready, updated with the latest data, can significantly enhance preparedness and response capabilities during future health emergencies.

\subsection{Simulation model requirements}
\label{s:simulation_model_requirements}
To accurately capture the complexities of vaccine manufacturing supply chains, the discrete-event simulation model must meet several key requirements, next to the main flow of goods (production processes):

\begin{itemize}
    \item \textbf{Integrated QA/QC module:} The model must include a dedicated QA/QC module that represents the personnel resources, different testing steps for quality control, sequential testing possibility, QA activities like document reviews and approvals, investigations of deviations and out of specification (OOS) results, and final batch release. 
    \item \textbf{Integrated raw materials module:} It must include a raw materials module that captures the procurement process, including supplier selection, lead time variability, and inventory management, to reflect the critical role of raw material availability in vaccine manufacturing.
    \item \textbf{Stochastic supply and process dynamics:} Procurement and production processes should be modeled with realistic lead time and processing time distributions (e.g., triangular, lognormal), capacity constraints, and inventory or resource allocation policies to reflect real-world uncertainty across materials, personnel, and machines.
    \item \textbf{Flexible scenario generation:} It must allow dynamic modification of model parameters during simulation runtime to represent diverse disruption scenarios (e.g., supplier failures, workforce shortages, machine breakdowns) without requiring model recompilation.
    \item \textbf{Parallel processing capabilities:} The model should support running hundreds of simulation replications across multiple scenarios, leveraging parallel processing and automated result aggregation for robust statistical analysis within a reasonable time frame.
\end{itemize}

\subsection{Simulation model description}
\label{s:simulation_model}
To address the modeling requirements outlined in Section~\ref{s:simulation_model_requirements}, a discrete-event simulation model was developed in the Julia programming language \cite{bezanson2017julia}. The choice of Julia was motivated by its computational performance, extensive statistical libraries, and ability to handle complex stochastic processes efficiently. Existing commercial simulation tools (e.g., Arena, AnyLogic) were evaluated but found inadequate for the specific requirements of vaccine manufacturing supply chain modeling, particularly in terms of flexible scenario generation, stochastic raw material dynamics, and integration with optimization routines.

\begin{figure}[h!]
    \centering
    \includegraphics[width=1\textwidth]{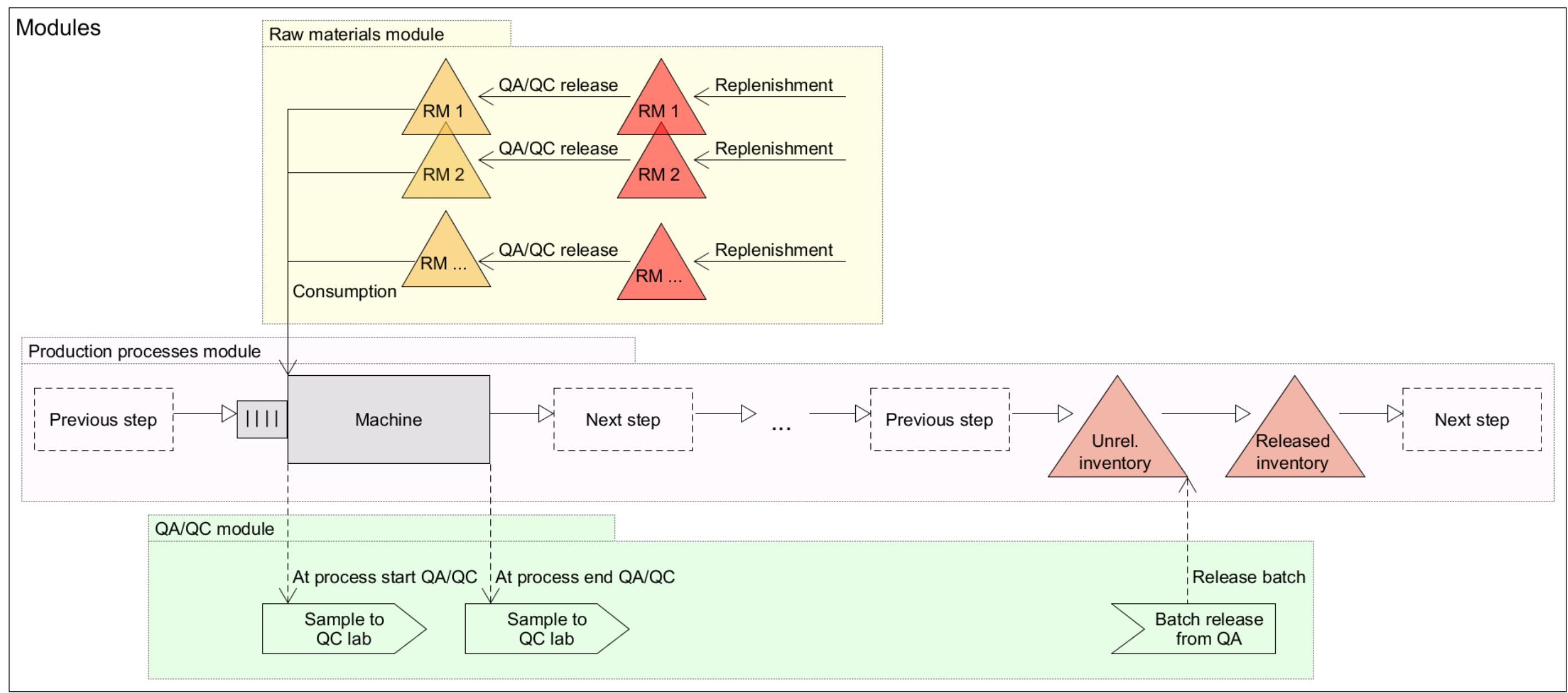}
    \caption{The relationship between the different modules in the simulation model. The production processes module is the main flow of goods, while the QA/QC and raw materials modules are inputs to and outputs from this main flow.}
    \label{fig:modules}
\end{figure}

The simulation model was developed through an iterative process involving close collaboration with vaccine manufacturing experts from CEPI, having expertise in vaccine supply chain management, manufacturing, analytical testing and quality assurance. This collaborative approach ensured that the model accurately reflects real-world manufacturing constraints, quality and regulatory requirements, and operational challenges. The model architecture follows a modular design with three interconnected components: (1) production processes, (2) QA/QC activities, and (3) raw materials procurement. This modular structure enables independent validation of each component while maintaining system-level integration. The production processes module is considered as the main flow of goods. The other modules (QA/QC and raw materials) are inputs and outputs of this main flow. Figure~\ref{fig:modules} provides an overview of the different modules and their interactions within the simulation model. 

The model implements a discrete-event simulation paradigm where the system state evolves through discrete time points corresponding to significant events (batch completions, resource availability changes, raw material deliveries). This approach naturally captures the batch-oriented nature of vaccine manufacturing while allowing for complex interactions between production stages, QA/QC activities, and supply chain dynamics. The simulation engine supports parallel execution across multiple scenarios and replications.

Key model features include dynamic resource allocation, stochastic process modeling, inventory management with expiration tracking, and event management for flexible disruption scenario implementation. The model generates detailed performance metrics at both aggregate and granular levels, supporting both strategic decision-making and operational optimization. Finally, the model generates automatically a visual overview of all the input data, automatic bottleneck detection and various plots to support analysis of the output data, which is combined in a report.

Model validation was conducted through multiple approaches: (1) unit testing of individual components against analytical solutions, (2) face validation with domain experts to ensure realistic system behavior, and (3) sensitivity analysis to verify model responsiveness to parameter changes. The model successfully reproduced key performance patterns observed in practice, including QA/QC bottlenecks during scale-up and inventory buildup during raw material disruptions.

The simulation model incorporates stochastic elements to capture the inherent uncertainty in vaccine manufacturing processes. Processing times and yield rates are modeled using probability distributions, reflecting real-world variability in production cycles. These probability distributions are also used for typical tasks and events of pharmaceutical manufacturing like document and batch record review, deviations, OOS results in analytical testing and their respective investigations.
Each time an event occurs, a new sample is drawn from the corresponding distribution to determine the actual processing time or yield for that specific instance. For each replication, the model uses a different random seed to ensure variability across simulation runs. For each scenario, the same seeds are used to ensure comparability between scenarios.

\subsubsection{Production processes}
\label{s:production_processes}

The production processes module operates at a user-defined granularity, allowing flexibility in modeling detail, from step-by-step representation to focusing on key interactions between major production stages. 
Rather than modeling technical process complexities (e.g., machine's physical dimensions, pressure conditions, temperature profiles), the model captures essential operational characteristics affecting supply chain performance: processing times, yield rates, capacity constraints, and resource requirements. 
This abstraction enables focus on system-level interactions and bottlenecks while maintaining applicability across diverse manufacturing configurations and technologies. 
The flexible design accommodates any production layout, making it adaptable to various vaccine manufacturing setups.

\subsubsection{Quality assurance and quality control activities}
\label{s:qa_qc_activities}
The QA/QC module models the complex testing and approval processes by capturing both the technical testing procedures (QC) and the approval processes (QA) that must be completed before batch release.

The QC testing process is modeled as a multi-stage activity where samples are collected at key production milestones (batch start, intermediate steps, batch completion) and processed through specialized laboratory teams. Each QC team represents a specific testing capability (e.g., microbiology, biochemistry) with dedicated personnel pools and equipment resources. The model explicitly tracks personnel constraints, as QA/QC specialists represent a limited and highly skilled resource that cannot be easily scaled during disruptions.
A special case of QC testing is in-process control (IPC) testing, which is performed during production to monitor critical quality attributes and ensure process consistency. In our model, IPC testing is performed by the production personnel and does not require additional QC personnel.

\begin{figure}[h!]
    \centering
    \includegraphics[width=1\textwidth]{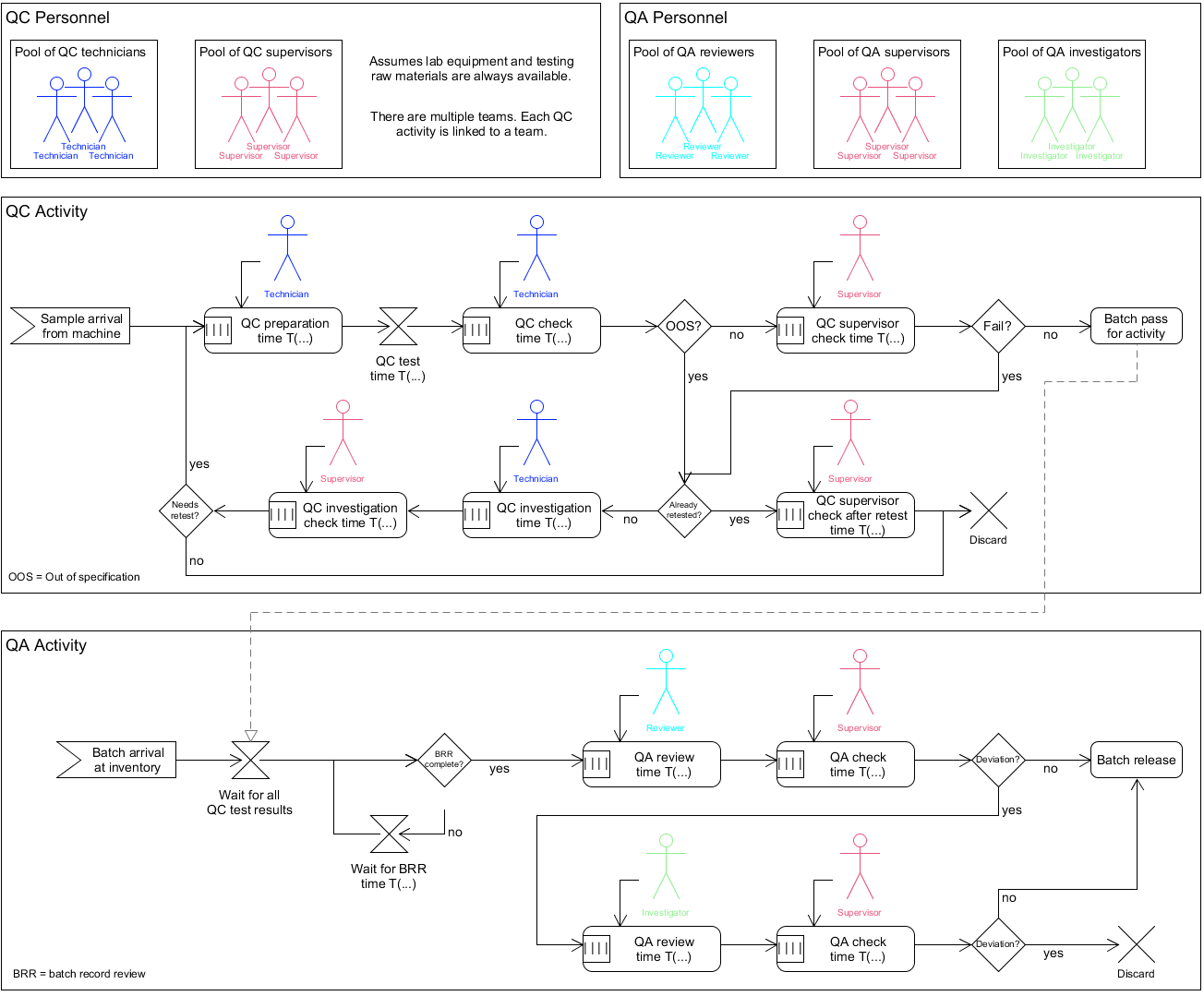}
    \caption{Overview of the QA and QC activity flows with personnel pools.}
    \label{fig:qaqc}
\end{figure}

Testing activities are modeled with stochastic processing times and failure probabilities. Failed tests trigger batch rejection and potential rework, creating additional delays and resource consumption. The model incorporates test interdependencies, where certain tests must be completed sequentially while others can proceed in parallel. QA review processes follow completion of manufacturing steps and analytical testing, with additional personnel requirements and approval timelines. Our model also reflects the use of QA/QC personnel for investigation of deviations and OOS results. An overview of the QA/QC activity flows is shown in Figure~\ref{fig:qaqc}.

Resource contention is explicitly modeled through finite personnel pools. When testing demand exceeds capacity, queues form and testing delays accumulate. The model tracks key metrics including personnel utilization rates, queue lengths, and testing throughput to identify QA/QC bottlenecks under different scenarios.

\subsubsection{Raw materials procurement and inventory management}
\label{s:raw_materials}
The raw materials module models the procurement and inventory management processes for critical manufacturing inputs. Raw materials encompass starting materials (e.g., DNA templates, host cells), direct production inputs (e.g., mRNA, lipids, adjuvants), indirect materials (e.g., vials, stoppers) and consumables (e.g., bioreactor bags, filters, pipette tips, cell culture media) required for vaccine manufacturing. 

For these critical raw materials, the model implements dynamic inventory policies that respond to production demand and supply uncertainty. Each raw material is characterized by consumption rates, procurement lead times, and storage constraints. The model supports multiple suppliers per raw material, allowing for diversification strategies and supplier switching during disruptions. To avoid unnecessary model complexity, only raw materials that manufacturers would classify as critical should be included.

\begin{figure}[h!]
    \centering
    \includegraphics[width=0.7\textwidth]{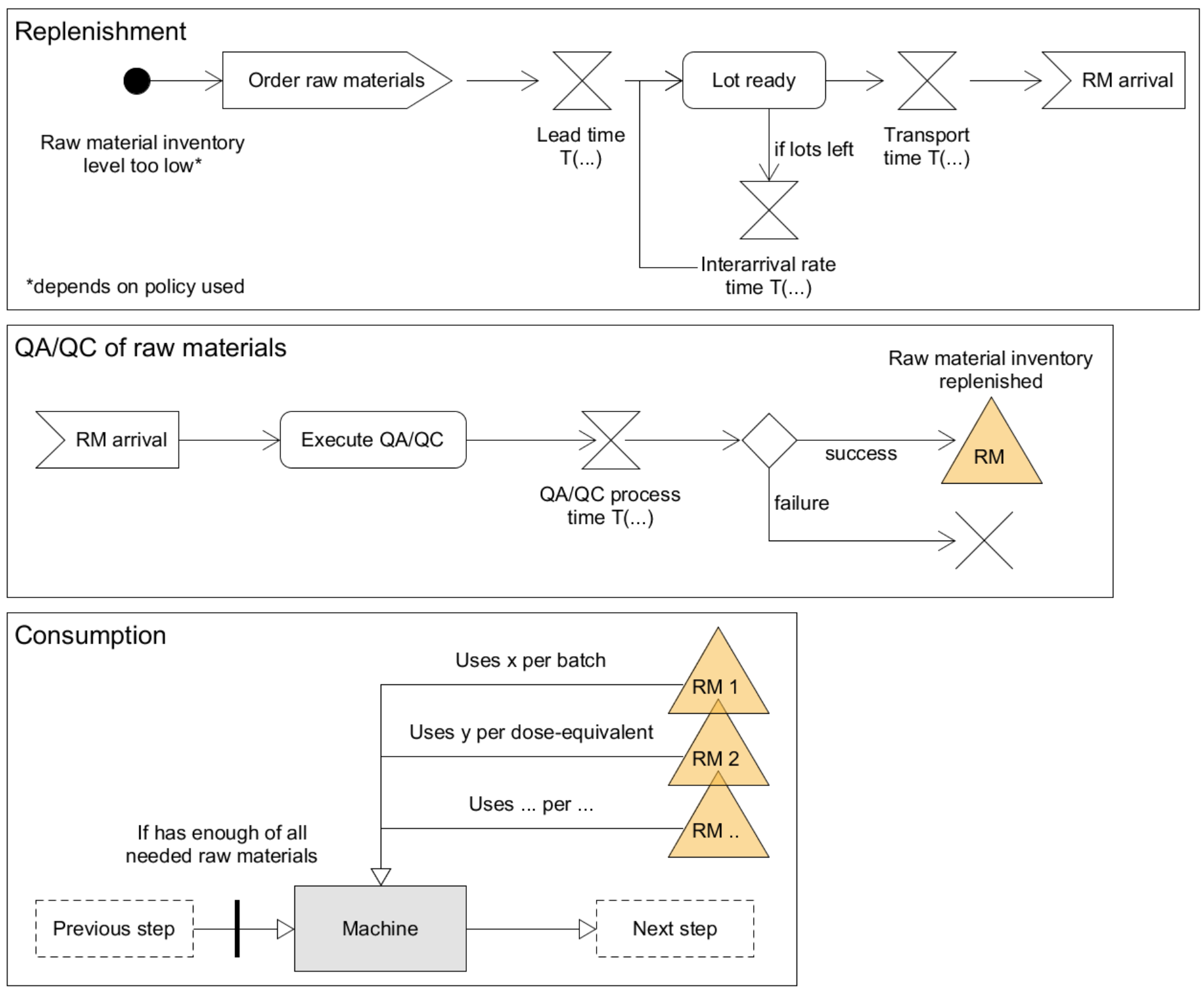}
    \caption{Overview of the raw materials replenishment, QA/QC and consumption.}
    \label{fig:raw_materials}
\end{figure}

Procurement processes are modeled as stochastic processes with lead time distributions, transportation times, lot sizes and interarrival times. During normal operations, suppliers operate according to their baseline performance parameters. During disruption scenarios, supplier availability and performance can be modified to reflect various supply chain interruptions. An overview of the raw materials replenishment, QA/QC and consumption is shown in Figure~\ref{fig:raw_materials}.

\subsection{Model assumptions}
The simulation model operates under several key assumptions that balance computational tractability with realistic representation of vaccine manufacturing operations, reflecting typical pandemic response manufacturing conditions. These assumptions enable the model to focus on critical interactions while maintaining computational efficiency for large-scale scenario analysis and have been discussed and validated with industry experts with significant experience in pandemic and regular (non-pandemic) vaccine manufacturing.

First, the production system operates as a push-based system where batches are initiated as soon as prerequisites are met: all raw materials are available, the next inventory has available capacity, and a machine is available. The model assumes continuous 24/7 production operations, with processing times scaled to account for actual operational hours when facilities operate with reduced shift patterns. For example, facilities having a 16-hour shift and 8 hours of downtime per day are modeled as an equivalent of a 24-hour shift to support continuous production operations. 
Second, processing times incorporate all necessary activities including setup, changeover, cleaning, and actual processing time. Personnel is not modeled explicitly in the production processes, as the focus is on machine constraints. 
Third, production at risk is permitted, enabling batches to advance through production stages before intermediate QA/QC clearance, though final batch release remains dependent on complete QA/QC release.
Fourth, facility operations assume no unplanned closures except for scheduled maintenance periods, such as end-of-year and summer holidays. QA/QC activities continue during production maintenance periods, allowing processing of accumulated samples. 
Fifth, cold storage systems maintain backup power during utility disruptions to preserve temperature-sensitive materials and intermediates.
Sixth, costs are not considered in this model, as the primary focus is on operational performance. The regulatory environment is assumed to remain constant throughout the simulation period, with QA/QC failure rates and investigation times following established probability distributions that can be updated during simulation to reflect real-time performance variations.
Seventh, when a batch fails retest, it is fully discarded and no salvage of materials is possible. Sequential testing dependencies are enforced where required, while parallel testing is utilized where operationally feasible. 
Eight, samples arriving at the QA/QC lab are processed with a priority system that takes into account the number of batches already present in the inventory where the batch release happens. For example, if there are already five batches waiting in an upstream inventory, a newly arrived batch from downstream with only two batches in the downstream inventory will be prioritized. If the number of batches in inventories is equal, the batch closest to final completion is prioritized. If there are still ties, the batch that arrived first is prioritized (first-in-first-out). QA/QC labs assume the availability of all necessary equipment and materials to perform testing without delays.
Ninth, expiries of raw materials, intermediate products nor final products are taken into account. Meaning not batches are lost due to expiry.
Finally, raw material suppliers operate independently with stochastic lead times, and the model assumes perfect information availability regarding inventory levels and material consumption. Transportation and logistics systems operate normally unless specifically disrupted in scenario configurations.

\subsection{Base case, scenarios and KPIs}
\label{s:base_case_scenarios_kpis}
When experimenting, the model defines the base case as a baseline scenario, representing normal operating conditions calibrated from industry data and expert input. Any scenario can then be designed by introducing specific changes or disruptions to stress test the system and evaluate its resilience. This flexible approach allows for systematic exploration of how different disruptions or policy interventions affect supply chain performance.

By comparing scenario outcomes to the base case, stakeholders can identify vulnerabilities, assess the impact of various disruptions, and evaluate the effectiveness of mitigation strategies. This enables policy makers and manufacturers to make informed decisions to improve the performance of their key performance indicators (KPIs), such as production speed, supply chain robustness, and resource utilization.

% ========================================================================================

\section{Numerical results: mRNA case study}
\label{s:case_study}
The methodology presented in Section~\ref{s:methodology} provides a general framework for analyzing vaccine supply chain disruptions. To demonstrate its validity, practical application and to generate actionable insights for pandemic preparedness, we apply this simulation modeling approach to a detailed case study of an mRNA vaccine manufacturing platform run in collaboration with CEPI. 

In recent years, messenger RNA (mRNA) vaccines have emerged as a transformative technology, offering rapid development timelines, high efficacy, and adaptable manufacturing processes \cite{Pardi2018}. Despite these advantages, mRNA vaccines present specific supply chain vulnerabilities that make them particularly susceptible to disruptions \cite{Rosa2021, Chaudhary2021}. For example, mRNA vaccines require specialized raw materials (such as specific nucleotides, capping reagents, enzymes, lipids) sourced from a limited number of suppliers, and stringent cold chain requirements throughout production and storage.

The mRNA case study examines the following research question:
\begin{quote}
    \textbf{RQ:} How do raw material procurement disruptions and QA/QC workforce constraints interact to affect mRNA vaccine manufacturing performance during pandemic response, and what could be an effective mitigation strategy?
\end{quote}

The following subsections detail the case study, the process of data collection and validation, define the key performance indicators (KPIs) used to assess supply chain performance, describe the base case, outline the disruption scenarios considered, and explain the experimental setup for the simulation study.

\subsection{Case study description}
\begin{figure}[h!]
    \centering
    \includegraphics[width=1\textwidth]{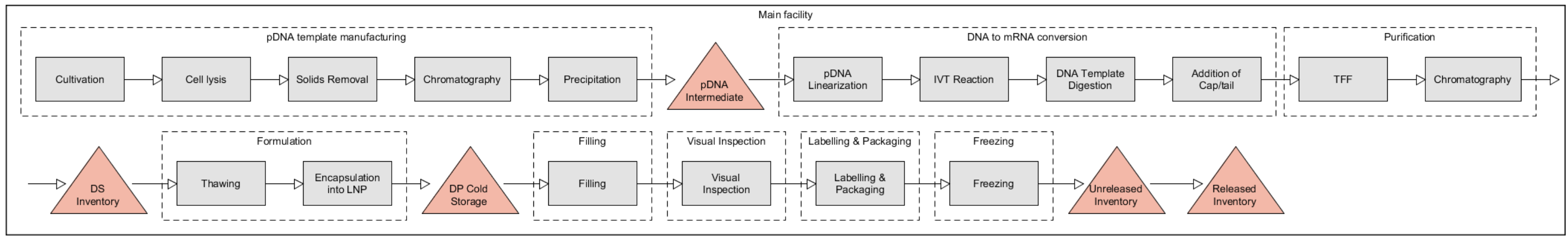}
    \caption{Product flow of the mRNA case study, including the different processes, intermediate inventories and final inventory.}
    \label{fig:product_flow}
\end{figure}

The case study is defined as a single, dedicated, facility producing tens of million doses of an mRNA vaccines per year. The facility has multiple processes from drug substance (DS) production up to drug product (DP) production, including QA/QC activities and raw material procurement. The diagram in Figure~\ref{fig:product_flow} provides an overview of the case study layout. As can be seen in the diagram, the manufacturing process modelled here starts with plasmid DNA production from E. coli cells, followed by RNA synthesis and purification, formulation filling, visual inspection, labelling and packaging. Our model assumes that the cell banks needed for plasmid DNA synthesis are already available. There are four intermediate inventories between the different processes, which are used to store the semifinished products and to buffer the production steps and a final inventory which contains the finished products. In total there are eighth different process steps, with in total 17 machines modelled. The complete overview of production times, capacities, yield rates, batch sizes, number of resources, maintenance periods, together with QA/QC activities data (preparation times, testing times, checking times, supervisory checks, failure probabilities, investigation times, personnel pools) and raw materials data (inital stockpiles, lead times, lot sizes, transportation times, quality control times for raw materials, rejection rates after quality control, replenishment policies, safety stocks, consumption rates, different suppliers splits (when multi supplier sourcing is used)) can be found in Appendix~\ref{s:appendix_complete_case_study_diagram}. Since the model is set up for a pandemic response scenario, the demand is considered infinite.

\subsubsection{Data collection and validation}
\label{s:data_collection}
To populate the model, data was collected through experts from CEPI with extensive experience in mRNA vaccine manufacturing, QA/QC and supply chain management. The experts have direct experience with pandemic-scale vaccine manufacturing and supply chain management from two pandemics: influenza H1N1 swine flu in 2009 and Covid-19 from 2020 onwards and are currently actively involved in pandemic preparedness initiatives.

The interviews followed a semi-structured format, organized over eight sessions lasting one to two hours each, and covered four main areas: (1) production process parameters including processing times, yield rates, deviation rates, and capacity constraints; (2) QA/QC activities including personnel requirements, testing times, failure probabilities; (3) raw material procurement including lead times, supplier characteristics, lot sizes, transportation times, and consumption; and (4) general parameters, including maintenance and simulation periods. Interviews were conducted jointly to have an agreed on (distribution of values for) parameters. Finally, disruption scenarios and mitigation strategies were co-created together with the two experts to ensure relevance and realism.

The collected data was validated through multiple approaches. First, parameter ranges were cross-validated between the two experts. Where discrepancies existed, conservative estimates were adopted. 
Second, for data that the experts did not have direct knowledge of or were unable to share, we used parameters found or calculated based on existing literature.
Finally, together with experts, one order-of-magnitude validation session was performed by comparing model outputs with known industry production scales and timelines.

\subsubsection{Key performance indicators}
\label{s:key_performance_indicators}
To evaluate the performance and resilience of the mRNA vaccine supply chain under various disruption scenarios, we define a set of key performance indicators (KPIs) that capture both operational efficiency and supply chain robustness. These KPIs are designed to provide actionable insights for manufacturers and policymakers by measuring critical aspects of vaccine manufacturing, from initial response capability to sustained output capacity.

The KPIs are categorized into three main areas: (1) responsiveness metrics that measure the speed of pandemic response, (2) throughput metrics that assess production capacity and efficiency, and (3) bottleneck analysis metrics.

\paragraph{Pandemic responsiveness metrics}
\begin{itemize}
    \item \textbf{Time to first dose:} The elapsed time from production initiation to the release of the first vaccine batch after complete QA/QC release. This metric is key for public health authorities to start vaccination campaigns. 
    \item \textbf{Time to 50 million doses:} The time required to produce and release 50 million vaccine doses, reflecting the system's ability to meet large-scale vaccination demands within a defined timeframe.
    \item \textbf{Cumulative doses after 365 days:} The total number of vaccine doses produced and released within the first year of production.
    \item \textbf{Supply chain recovery time:} The duration in weeks required for the system to return to normal operations, measuring the system's ability to bounce back from shocks. This is calculated by comparing the scenario's daily values of a smoothed 30-day rolling average to the base case daily values. Once a daily value is significantly lower ($p<0.05$) than the base case value, the scenario enters its disruption period and end when the daily value is no longer significantly lower. 
\end{itemize}

\paragraph{Throughput metrics}
\begin{itemize}
    \item \textbf{Maximum production capacity:} The monthly and cumulative production capacity achieved after the first dose has been released.
    \item \textbf{Batch lead time distribution:} The distribution of the durations from batch initiation to final QA/QC release, encompassing all production, testing, and approval processes.
\end{itemize}

\paragraph{Bottleneck analysis metrics}
\begin{itemize}
    \item \textbf{Machine utilization rates:} The average utilization rates of production machines, measured as the percentage of available capacity used over time.
    \item \textbf{QA/QC personnel utilization rates:} The average utilization rates of critical QA/QC personnel pools (technicians, supervisors, reviewers), measured as percentage of available capacity.
    \item \textbf{Raw material stockout frequency:} The number of stockout occurrences for each critical raw material, where production is delayed due to inventory unavailability.
\end{itemize}

\subsubsection{Base case and scenarios}
\label{s:scenarios}
The base case represents normal mRNA vaccine manufacturing operations during a pandemic, where production is initiated to meet a target of 50 million doses (50$\mu$g of active pharmaceutical ingredient per dose) within the first year, reflecting typical order-of-magnitude requirements specified for instance in HERA contracts. The base case assumes an initial stockpile for the raw materials equivalent to 3 to 6 months (depending on the raw material) of production capacity. Raw material warehouse storage capacity is assumed to be unconstrainted.

The scenarios in this paper were selected to represent different disruptions. Scenarios 1 to 5 test system vulnerabilities under stress, while scenario 6 evaluates a potential resilience investment. The combination allows for assessment of supply chain weak points and the effectiveness of countermeasures.
These scenarios were generated based on interviews with field experts and should reflect real-life situations, backed-up by evidence reported in literature. The scenarios are defined as follows (the affected components and data points can be found in Table~\ref{tab:scenario_overview}):
\begin{itemize}
    \item[S1] \textbf{Production facility shutdown (complete disruption)}
    A complete four-week shutdown of the production facility, representing severe quality incidents, machine failures, or regulatory suspensions that require full facility remediation. This scenario tests the system's ability to recover from complete production halts.
    Rationale: Major quality issues or machine failures can force complete facility closures, as observed during COVID-19 \cite{Manas2021}.

    \item[S2] \textbf{Workforce capacity reduction (capacity disruption)} 
    A four-week period of reduced operational capacity, with machine capacity reduced to 50\% and QA/QC personnel halved, representing pandemic-related illness, quarantine measures, or workforce shortages.
    Rationale: Disease outbreaks, quarantine requirements, and workforce shortages were common during COVID-19 \cite{Smeaton2021}.
    
    \item[S3] \textbf{Critical raw material unavailability (supply disruption)}
    Complete unavailability of enzymes for three months, representing supplier failures, transportation disruptions, or raw material shortage scenarios that completely eliminate access to critical inputs.
    Rationale: Single-source suppliers for specialized materials create vulnerabilities to complete supply interruptions \cite{Blome2009}.
    
    \item[S4] \textbf{Power outage (infrastructure disruption)} 
    A 24-hour power outage resulting in loss of all work-in-progress batches and all QA/QC samples, while batches in cold storage remain protected by backup power systems.
    Rationale: Infrastructure failures can cause catastrophic loss of in-process materials \cite{Kelly2015}.
    
    \item[S5] \textbf{Raw material procurement lead time increase (chronic supply stress)}
    Permanent increase in lead times for vials and filters throughout the simulation period, representing long-term supplier capacity constraints or transportation disruptions.
    Rationale: Pandemic demand surges create sustained supply chain stress and extended lead times for specialized materials \cite{Feddema2023}.
    
    \item[S6] \textbf{QA/QC personnel capacity increase (mitigation strategy)}
    Doubling of all QA/QC personnel throughout the simulation period, representing proactive capacity investments, overtime policies, or emergency hiring and training programs.
    Rationale: Industry experts identified QA/QC as a critical bottleneck; this scenario evaluates the effectiveness of QA/QC capacity investments as a mitigation strategy.
\end{itemize}

\begin{table}[h!]
    \centering
    \caption{Scenario overview: impacted variables and value changes}
    \resizebox{\textwidth}{!}{
    \begin{tabularx}{\textwidth}{>{\hsize=1\hsize}X >{\hsize=2\hsize}X >{\hsize=0.48\hsize}X >{\hsize=0.52\hsize}X}
        \toprule
        \textbf{Scenario} & \textbf{Affected Components} & \textbf{Baseline Value} & \textbf{Disruption Value} \\
        \midrule
        \addlinespace[0.5em]
        S1 - Shutdown & All machine states between 01/09/25 and 28/09/25 & Operating & Closed \\
        \addlinespace[0.5em]
        S2 - Disease outbreak & Reduced capacity between 01/05/25 and 28/05/25: & & \\ 
        & \hspace*{0.5em} QC technicians (all teams) \& QA reviewers \& QA supervisors & 2 & 1 \\
        & \hspace*{0.5em} QC supervisors (all teams) \& QA investigators & 1 & 0 \\
        & \hspace*{0.5em} Processing time (all machines) & baseline levels & $2\times$ baseline \\
        \addlinespace[0.5em]
        S3 - Raw material (RM) unavailability & Enzymes availability between 01/09/25 and 01/12/25 & Yes & No \\
        \addlinespace[0.5em]
        S4 - Power outage & Event on 01/09/25, all WIP and QC samples & Normal & Complete loss \\
        \addlinespace[0.5em]
        S5 - Raw material procurement lead time increase & Vials lead time (in weeks) & T(6,8,12) & T(24,32,48) \\
        & Filters lead time (in months) & T(1,2,3) & T(6,7,8) \\
        \addlinespace[0.5em]
        S6 - QA/QC personnel increase & All QC personnel & baseline levels & $2\times$ baseline \\
        & All QA personnel & baseline levels & $2\times$ baseline \\
        \bottomrule
        \multicolumn{4}{l}{\footnotesize Note: Values for lead times are given as triangular distributions T(min, mode, max).} 
        \end{tabularx}
        }
    \label{tab:scenario_overview}
\end{table}

\subsubsection{Statistical analysis plan and experiment setup}
Each scenario is run 100 times with different random number streams to ensure statistical robustness. Results are compared to the base case using appropriate statistical tests, with confidence intervals reported for key performance metrics. 

The simulation starts on 01/04/2025 and runs until 31/03/2028 (3 years) to allow the system to stabilize and to capture the effects of the disruptions. Quarter 2 is used as starting point in order to not begin during a maintenance period (end of year holidays). 

The simulation is run on a desktop computer with a 8-core CPU and 32 GB of RAM. Each replication takes approximately 5 minutes to run, resulting in a total simulation time of approximately 8 hours for all 100 replications per scenario. The simulation is run in parallel to speed up the process, with each replication running on a separate core.
The results of the simulation are saved in a SQLite database to be retrieved and create the figures using the Julia package Plots.jl.

% ========================================================================================

\subsection{Results}
\label{s:results}
This section presents the results of the mRNA vaccine supply chain case study, focusing on the base case and the impact of various disruption scenarios on key performance indicators (KPIs). The results are organized as follows: first, we present the base case analysis, followed by a selection of results comparing the base case to the most interesting scenarios. Finally, we discuss the implications of these findings for vaccine manufacturing resilience and pandemic preparedness.

\subsubsection{Base case analysis}
\label{s:base_case_analysis}
In this section, the results of the base case are presented. We start with the pandemic responsiveness metrics, followed by the throughput metrics, and end with the bottleneck analysis.

\begin{figure}[h!]
    \centering
    \includegraphics[width=0.7\textwidth]{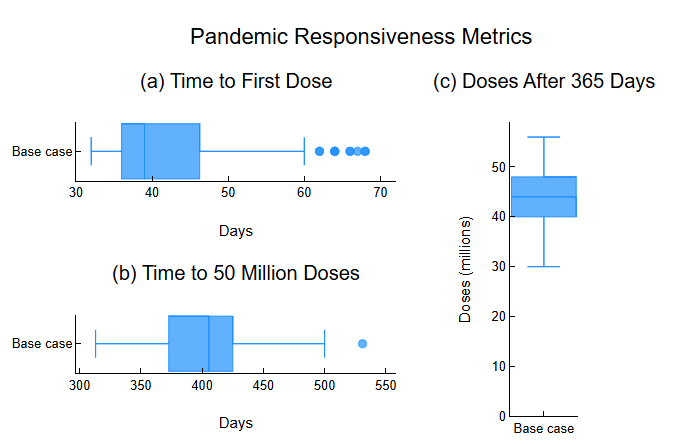}
    \caption{Base case pandemic responsive metrics: (a) time to first dose, (b) time to 50 million doses, and (c) number of doses after 365 days.}
    \label{fig:combinedTimeTo_bc}
\end{figure}

\newparagraph{Pandemic responsiveness metrics}
The first KPI analyzed is the time to first dose, which is a measure of how quickly a vaccine can be produced and released for distribution. 
Figure~\ref{fig:combinedTimeTo_bc}\hyperref[fig:combinedTimeTo_bc]{a} shows that the median of the time to first dose of the base case is 39 days, with 50\% of the replications having a result between 36 and 46 days, and a maximum of 68 days. 
These results indicate that, once a vaccine has been successfully developed and approved for production, the majority of simulation runs achieve first dose production within two months. 

The second KPI analyzed is the time to 50 million doses, which is a measure of how quickly a large quantity of vaccines can be produced and released for distribution. Figure~\ref{fig:combinedTimeTo_bc}\hyperref[fig:combinedTimeTo_bc]{b} shows that the median of the time to 50 million doses of the base case is 406 days ($\sim$13.5 months), with 50\% of the replications having a result between 373 and 425 days, and an outlier of 531 days. This means that, after the first dose has been produced, it takes roughly 12 additional months to produce 50 million more doses. With a spread of over 6 months in time to 50 million doses, there is considerable variability in when this target will be achieved. 
This finding could inform health agencies to setup and plan their vaccination campaigns.

The third KPI looks at the number of doses produced after one year. The base case produces a median of 44 million doses after one year, with 50\% of the replications having a result between 40 and 48 million doses, a minimum of 30 million and a maximum of 56 million doses (Figure~\ref{fig:combinedTimeTo_bc}\hyperref[fig:combinedTimeTo_bc]{c}). 

\begin{figure}[h!]
    \centering
    \includegraphics[width=1\textwidth]{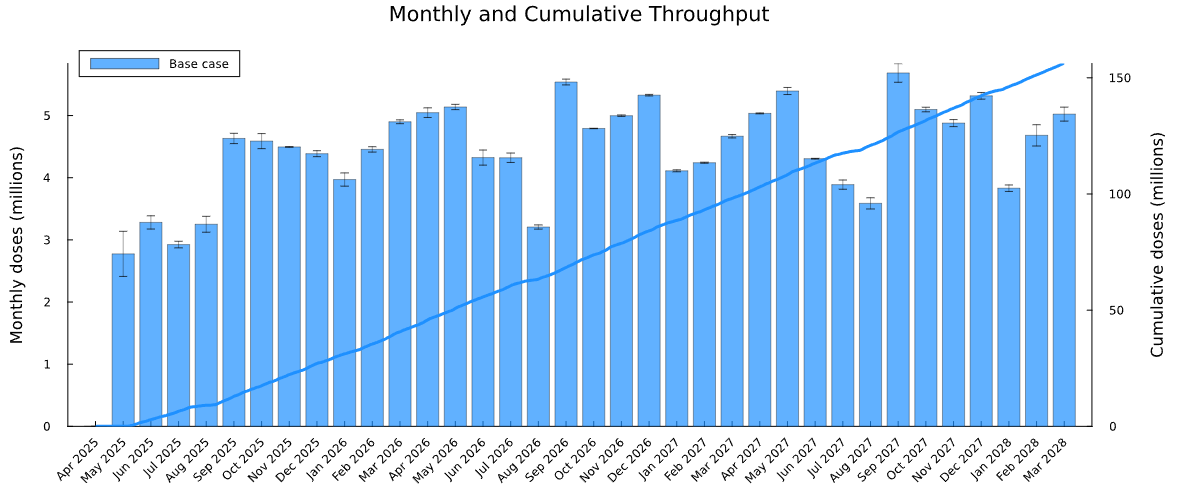}
    \caption{Monthly and cumulative throughput of the base case.}
    \label{fig:combinedThroughput_bc}
\end{figure}

\newparagraph{Throughput metrics}
Figure~\ref{fig:combinedThroughput_bc} shows the monthly and cumulative throughput of released doses of the base case. As can be seen, in April 2025, no doses are produced and the subsequent three four months, around 3 million doses are produced, which is due to the warm-up period of the production process. 
After this warm-up period, the production process stabilizes. After the first dose is produced an average of 4.43 million doses are produced per month (over the three years). The figure shows that the production fluctuates between four and five million doses per month (this is a difference of only two batches per month) and shows clearer drops during summer maintenance. The cumulative number of doses is shown on the line graph, which shows that after three years, 156 million doses have been produced.

\begin{figure}[h!]
    \centering
    \includegraphics[width=1\textwidth]{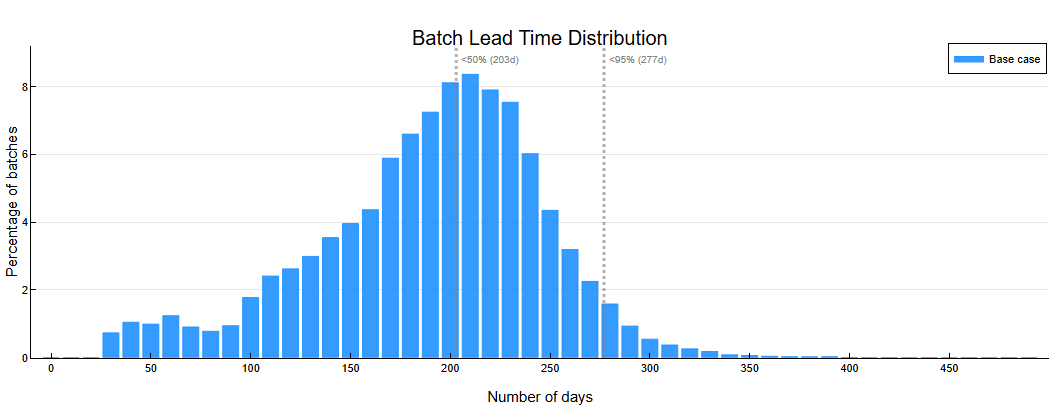}
    \caption{Batch lead time distribution of the base case (in bins of 10 days).}
    \label{fig:batchLeadTimeBar_bc}
\end{figure}

The next KPI analyzed is the batch lead time distribution, which is a measure of how long it takes to produce a batch of vaccine when it enters the first machine to batch release at the final inventory. The batch lead time distribution (Figure~\ref{fig:batchLeadTimeBar_bc}) is grouped in bins of 10 days to reduce noise.
At the start, from 30 to 90 days, we see an equal percentage of batches finishing between that time period, indicating there were no major bottlenecks at QA/QC (due to queue buildup or sample retesting) or occasional raw material stockout.
From 100 days onwards, we see a clear increase in batch lead time, with a peak around 210 days. This is mainly due to a major queue buildup at QA/QC. The tail that goes over 400 days shows batches undergoing many retests over the course of their production. The two vertical lines indicate the average batch lead time at 203 days and the 95th percentile at 277 days.

\begin{figure}[h!]
    \centering
    \includegraphics[width=1\textwidth]{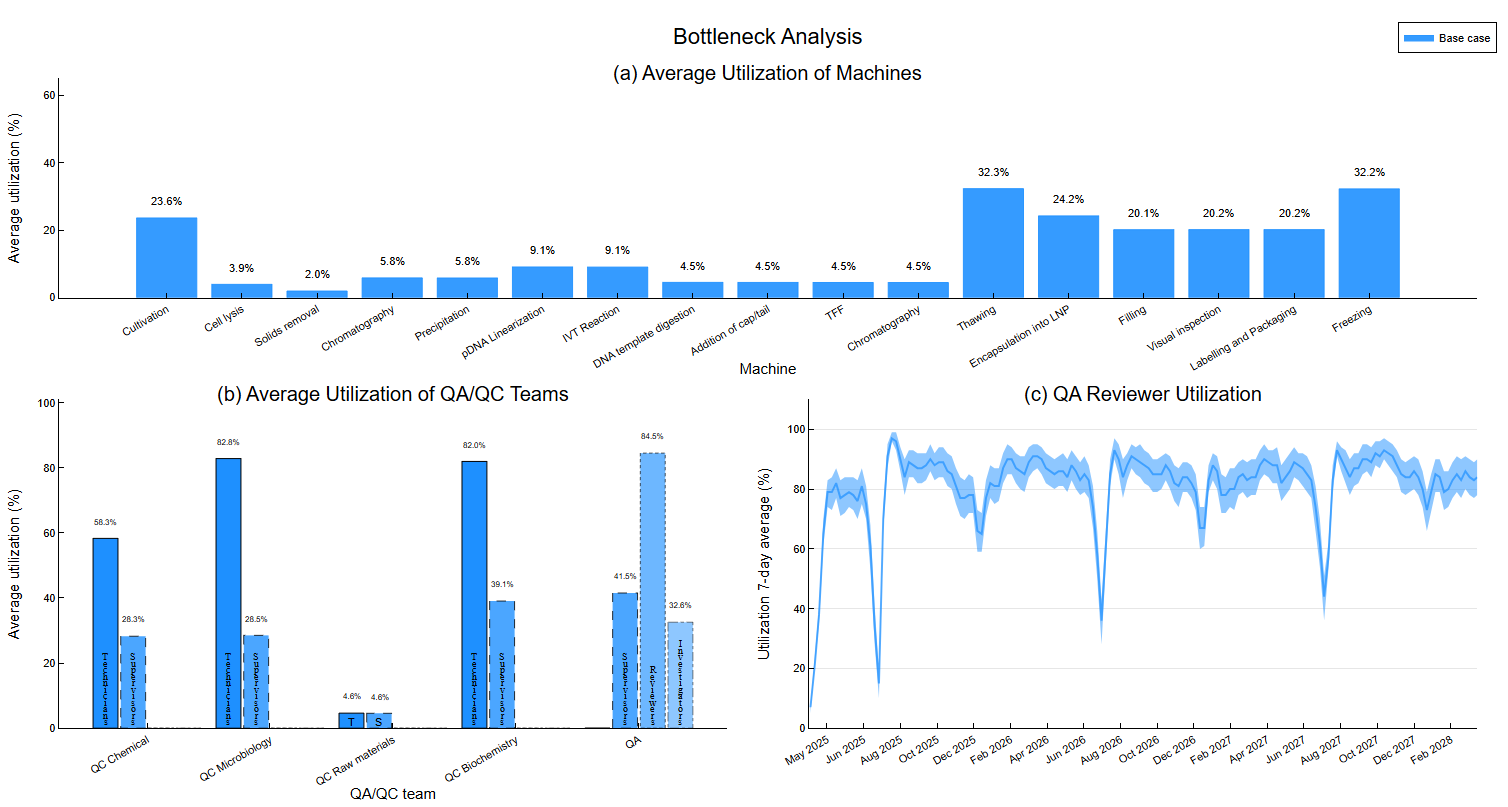}
    \caption{Base case bottleneck analysis: (a) Average machine utilization, (b) Average QA/QC personnel utilization by team and category (T=Technicians, S=Supervisors, R=Reviewers, I=Investigators), and (c) QA Reviewers utilization over time.}
    \label{fig:bottleneckAnalysis_bc}
\end{figure}

\newparagraph{Bottleneck analysis metrics}
To systematically identify the factors constraining production performance, we analyze potential bottlenecks across three critical areas: machine utilization, QA/QC personnel capacity, and raw material availability. 
Figure~\ref{fig:bottleneckAnalysis_bc}\hyperref[fig:bottleneckAnalysis_bc]{a} shows that machines are not operating at capacity, with the highest utilization rate being only 32.3\% for the thawing machine, indicating that production machines have substantial unused capacity available. 
However, analysis of QA/QC personnel utilization in Figure~\ref{fig:bottleneckAnalysis_bc}\hyperref[fig:bottleneckAnalysis_bc]{b} reveals a markedly different pattern, with personnel utilization rates reaching 84.5\% for QA reviewers, indicating near-maximum capacity utilization. 

When looking at detailed utilization rates of the QA reviewers, Figure~\ref{fig:bottleneckAnalysis_bc}\hyperref[fig:bottleneckAnalysis_bc]{c} shows that the QA/QC personnel are heavily utilized and are often close to their maximum capacity. This leads to a queue buildup, which causes delays in the QA/QC activities and, consequently, in the production of the vaccine. We do see that during the maintenance period the utilization goes down when the queue has been cleared. 

Raw material availability, as can be found in Appendix~\ref{s:appendix_additional_results} Table~\ref{tab:raw_material_stockouts}, demonstrates minimal stockout occurrences under base case conditions, with only occasional shortages (less than a day per year on average) that does not significantly constrain overall production. 
Figure~\ref{fig:rawMaterials_bc_s5_s6} shows the inventory levels of two critical raw materials, chromatography gels (used in the first process) and filters (used in the fifth process), over the three year simulation period. The raw materials inventory levels are represented in batch equivalents, to be able to compare them. One batch equivalent is the raw material's consumption for one batch going through the machine where the raw material is needed.
The figure shows that the inventory level of the chromatography gels drops straight at the start of the simulation, while it warms up. Once the system reaches its steady-state, it builds up again an inventory of around 30 to 40 batch equivalents. Filters show a slower decline at the start, but clearly show a seasonal pattern, with inventory levels increasing during the maintenance periods. In both cases, the inventory levels remain high and never reach zero, indicating that there are no stockouts of these raw materials in the base case.

\subsubsection{Scenario analysis}
\label{s:scenario_analysis}
In this section, the results of the scenario analysis are presented. The scenarios are compared to the base case to see the impact of the disruptions on the mRNA vaccine manufacturing. To reduce visual clutter, for some comparisons, a selection of scenarios is made. 

\begin{figure}[h!]
    \centering
    \includegraphics[width=1\textwidth]{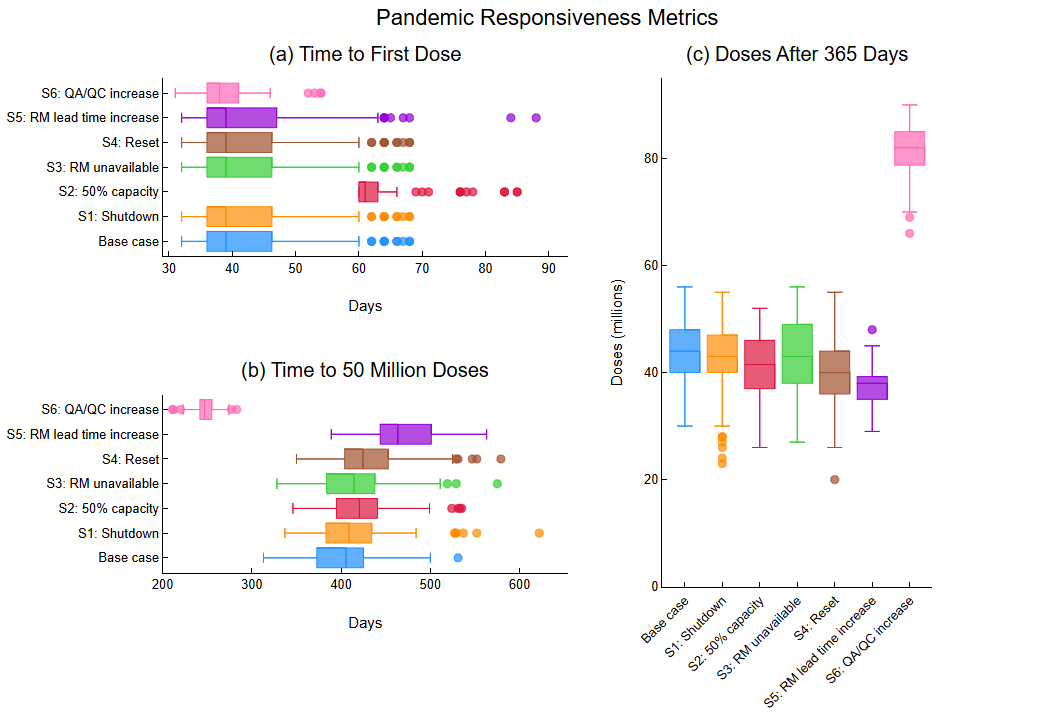}
    \caption{Pandemic responsive metrics for all scenarios: (a) time to first dose, (b) time to 50 million doses, and (c) number of doses after 365 days.}
    \label{fig:combinedTt1dDoses}
\end{figure}

\newparagraph{Pandemic responsiveness metrics}
Figure~\ref{fig:combinedTt1dDoses}\hyperref[fig:combinedTt1dDoses]{a} shows the time to first dose across different scenarios. Compared to the base case, scenarios 1, 3, and 4 demonstrate no difference, as the disruption occurs only after the first dose is produced. Scenario 5 shows only a slight variation, due to the stochastic behavior of the model. However, scenario 2 exhibits a clear delay with a median value of 61 days due to reduced capacity early in the simulation period. Conversely, scenario 6 shows improved performance with a narrower distribution, indicating greater certainty in achieving faster dose release through increased QA/QC capacity. While 75\% of replications over all scenarios achieve first dose within 46 days or less (except scenario 2), extreme values extending up to 3 months still occur.

Figure~\ref{fig:combinedTt1dDoses}\hyperref[fig:combinedTt1dDoses]{b} and Figure~\ref{fig:combinedTt1dDoses}\hyperref[fig:combinedTt1dDoses]{c} show the time to produce 50 million doses and the number of doses produced after 365 days across different scenarios, respectively. 
Figure~\ref{fig:combinedTt1dDoses}\hyperref[fig:combinedTt1dDoses]{b} shows for scenario 6 a clear time reduction for the number of days it takes to reach 50 million doses. Scenario 5 shows extended time, other scenarios show (slightly) longer times. 
In Figure~\ref{fig:combinedTt1dDoses}\hyperref[fig:combinedTt1dDoses]{c}, except for scenario 6, all scenarios demonstrate decreased production compared to the base case. Table~\ref{tab:scenario_doses_comparison} provides detailed statistical comparisons, revealing that in the first 12 months scenarios 2, 4, 5, and 6 show significant differences ($p<0.05$), while scenarios 1 and 3 do not differ significantly from baseline performance. Scenario~6 achieves the most substantial improvement (+84.8\%, $p<0.001$), demonstrating that increased QA/QC capacity represents an effective mitigation strategy. However, the improvement is non-linear, doubling QA/QC personnel does not double vaccine output, because other system constraints, including machine capacity and raw material availability, ultimately limit production scaling.

\begin{table}[h!]
    \centering
    \caption{Scenario comparison: dose production (millions) with 95\% confidence intervals, percentage change from base case, and statistical significance (t-test p-values, significant values ($p<0.05$) are highlighted in bold) for 12-month and 36-month periods}
    \resizebox{\textwidth}{!}{
        \begin{tabular}{lrrrrrr}
            \toprule
            \multirow{2}{*}{\textbf{Scenario}} & \multicolumn{3}{c}{\textbf{12 Months}} & \multicolumn{3}{c}{\textbf{36 Months}} \\
            \cmidrule(lr){2-4} \cmidrule(lr){5-7}
            & \textbf{Doses (M)} & \textbf{$\Delta$} & \textbf{$p$-value} & \textbf{Doses (M)} & \textbf{$\Delta$} & \textbf{$p$-value} \\
            \midrule
            Base case & 43.8 [42.7; 44.9] & — & — & 156.1 [154.0; 158.2] & — & — \\
            S1: Shutdown & 42.9 [41.6; 44.1] & -2.1\% & 0.261 & 154.5 [152.3; 156.6] & -1.0\% & 0.270 \\
            S2: 50\% capacity & 41.5 [40.4; 42.6] & -5.3\% & \textbf{0.003} & 150.9 [148.4; 153.3] & -3.3\% & \textbf{0.001} \\
            S3: RM unavailable & 43.0 [41.8; 44.3] & -1.9\% & 0.365 & 153.4 [151.2; 155.7] & -1.7\% & 0.086 \\
            S4: Reset & 39.6 [38.4; 40.7] & -9.6\% & \textbf{$<$0.001} & 150.6 [148.4; 152.8] & -3.5\% & \textbf{$<$0.001} \\
            S5: RM lead time increase & 37.4 [36.7; 38.1] & -14.7\% & \textbf{$<$0.001} & 125.5 [124.5; 126.5] & -19.6\% & \textbf{$<$0.001} \\
            S6: QA/QC increase & 81.0 [80.1; 82.0] & +84.8\% & \textbf{$<$0.001} & 279.6 [277.7; 281.5] & +79.1\% & \textbf{$<$0.001} \\
            \bottomrule
        \end{tabular}%
    }
    \label{tab:scenario_doses_comparison}
\end{table}

\begin{figure}[h!]
    \centering
    \includegraphics[width=1\textwidth]{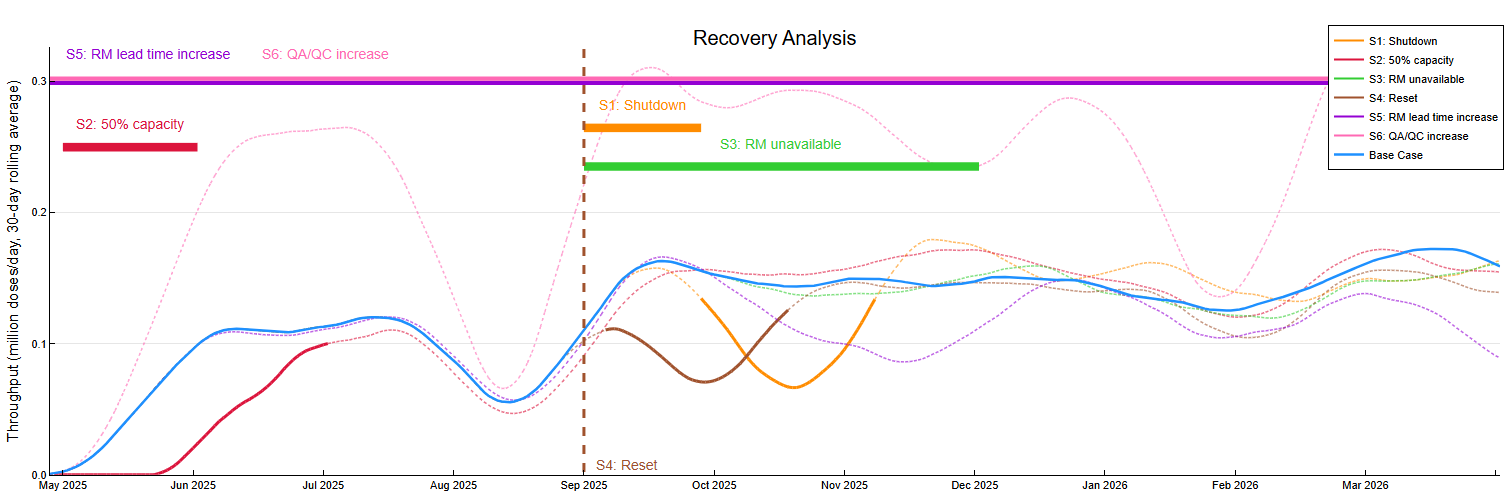}
    \caption{Supply chain recovery analysis: production rates (smoothed 30-day rolling average) compared to base case performance across all scenarios. Values highlighted show the recovery period after a disruption has occurred. Only significantly lower ($p<0.05$) values compared to the base case are highlighted. Continuous disruptions (scenarios 5 and 6) are not highlighted.}
    \label{fig:recovery}
\end{figure}

The final pandemic responsiveness metric analyzed is the supply chain recovery time, which measures how quickly the system can return to near-normal operations after a disruption. Figure~\ref{fig:recovery} shows the production rates compared to the base case with a smoothed 30-day rolling average for all scenarios. Only the base case and values of scenarios experiencing a disruption are highlighted. The moment the disruption occurs is indicated with a dashed vertical line for sudden disruptions (scenario 4) and with bars for the other scenarios. Depending on the scenario, different dynamics can be observed. Continuous disruptions (scenarios 5 and 6) are not highlighted.

\begin{figure}[h!]
    \centering
    \includegraphics[width=1\textwidth]{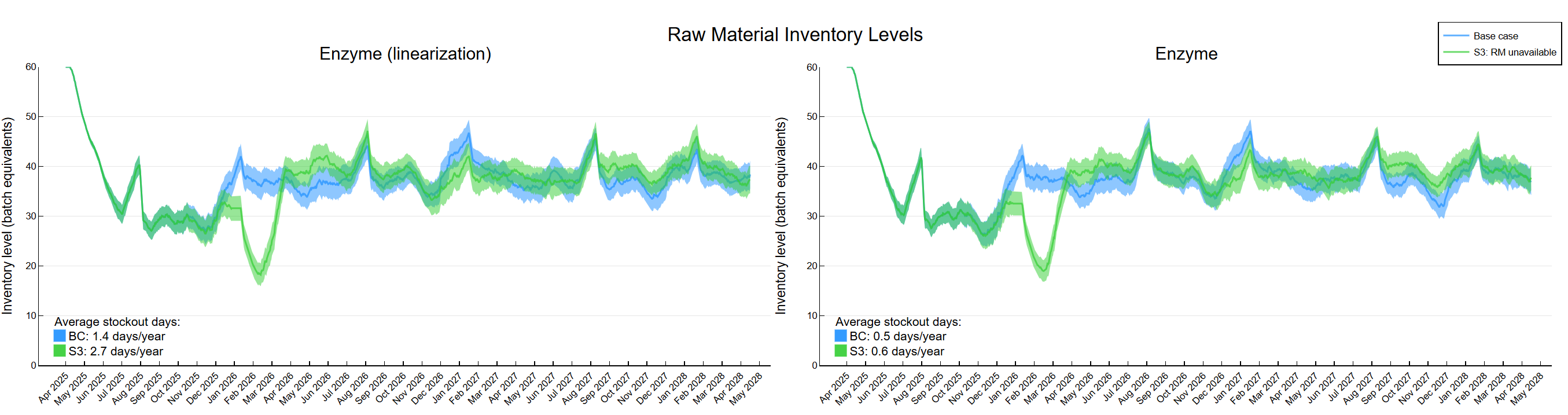}
    \caption{Inventory levels of the two types of enzymes (raw materials) of the base case and scenario~3.}
    \label{fig:rawMaterials_bc_s3}
\end{figure}

Scenario 1 (shutdown) shows a delayed drop in production, as QA/QC can still operate and release batches that were already finished. Only after three weeks, the output drops below the threshold. This scenario needs 6 weeks to recover.
Scenario 2 (reduced capacity) shows an immediate drop in production, as the capacity is reduced from the start. This scenario needs 12 weeks to recover.
Scenario 3 (unavailability of enzymes) shows no significant impact on production, as the initial stockpile of enzymes is in most of the replications sufficient to cover production during the disruption (Figure~\ref{fig:rawMaterials_bc_s3}). This scenario needs no time to recover.
Scenario 4 (reset of all WIP and QC samples) shows a drop in production when the disruption occurs, as all WIP and QC samples are lost. Because Figure~\ref{fig:recovery} uses a rolling average to smooth out the results, the drop is not sharp. This scenario needs 9 weeks to recover.
Scenario 5 (increased lead times of raw materials) shows a continuous drop in production after the initial stockpile of raw materials is used up. After that the longer lead times cause stockouts throughout the simulation period. Even though the production sometimes goes above the threshold, we still consider this scenario not recovered, as the production is not stable. 
Scenario 6 (increased QA/QC capacity) shows no negative impact on production and is always above the threshold. 

\begin{figure}[h!]
    \centering
    \includegraphics[width=1\textwidth]{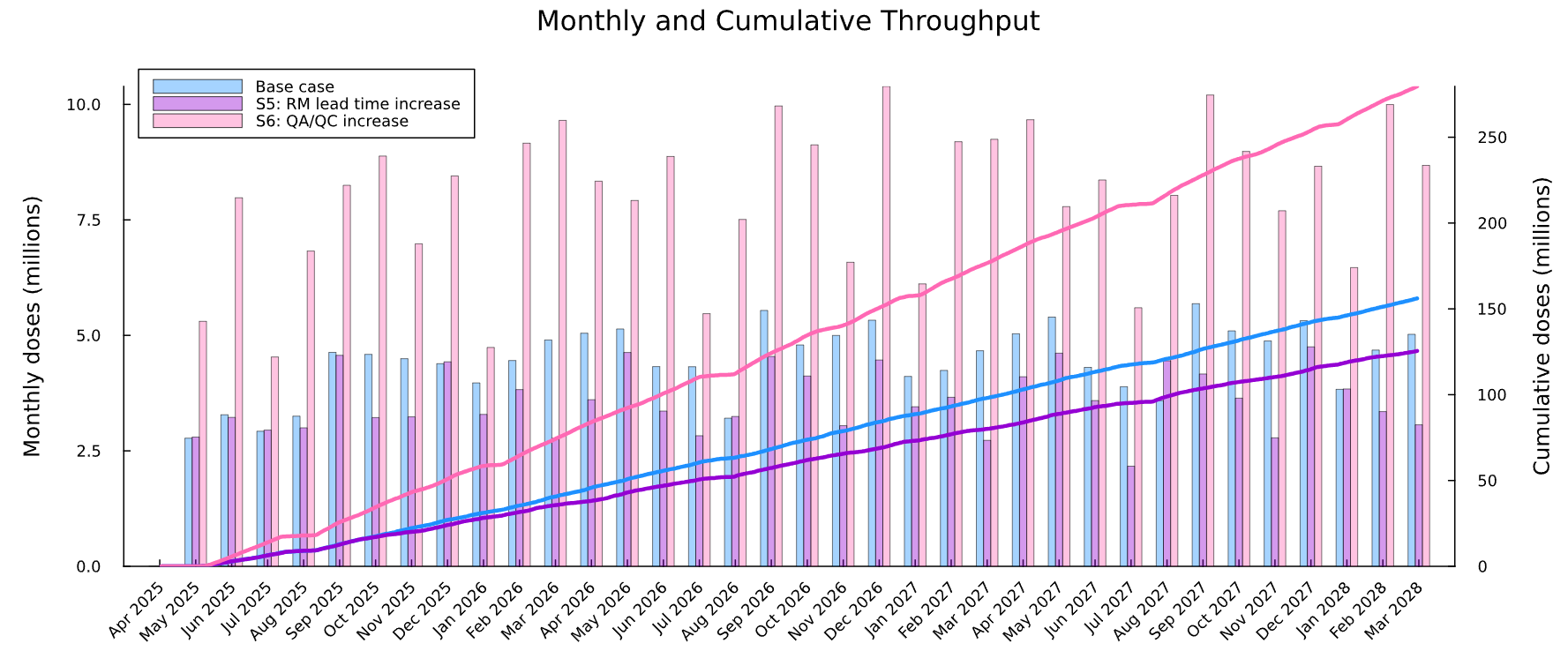}
    \caption{Monthly and cumulative vaccine throughput of the base case, scenario 5 and scenario 6.}
    \label{fig:combinedThroughput_bc_s5_s6}
\end{figure}

\newparagraph{Throughput metrics}
At full production capacity (Table~\ref{tab:scenario_doses_comparison}), the base case produces 156.1 million doses over three years. Scenario 6, with doubled QA/QC personnel, achieves a significant increase to 279.6 million doses (+79.1\%, $p<0.001$), while scenario 5, with increased raw material lead times, drops to 125.5 million doses (-19.6\%, $p<0.001$). Scenarios 2 and 4 show a significant small reduction (around $-3\%$). Scenarios 1 and 3 show minor reductions without statistical significance. 
Figure~\ref{fig:combinedThroughput_bc_s5_s6} shows the monthly and cumulative throughput of the base case, scenario 5 and scenario 6. Scenario 5 follows the base case closely for the first six months and then starts to diverge, producing significantly less doses. This is due to the initial stockpile of raw materials being used up and the longer lead times causing stockouts. Scenario 6 shows a significant increase in production from the first production month onwards. Only during the maintenance periods, the output is clearly lower.

\begin{figure}[h!]
    \centering
    \includegraphics[width=1\textwidth]{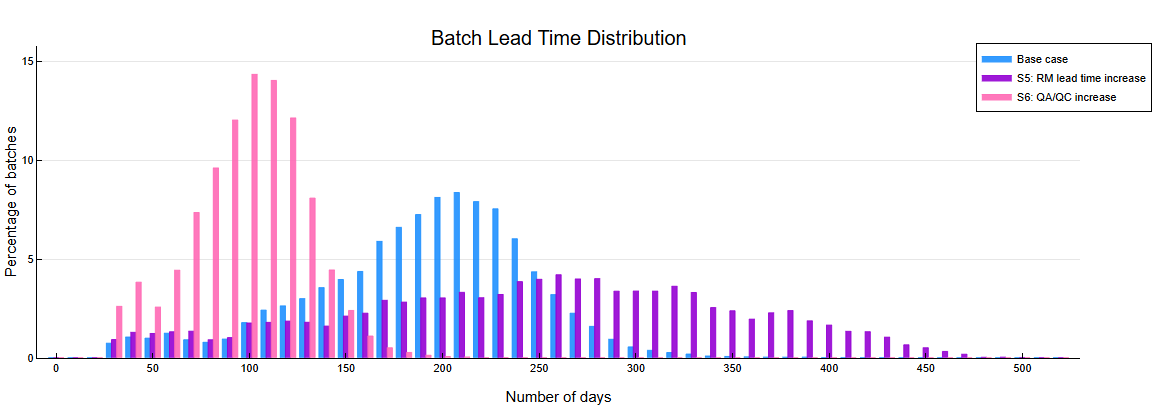}
    \caption{Batch lead time distribution of the base case and scenarios 5 and 6.}
    \label{fig:batchLeadTimeBar_bc_s5_s6}
\end{figure}

In Figure~\ref{fig:batchLeadTimeBar_bc_s5_s6}, the batch lead times are compared across the base case and scenarios 5 and 6. Scenario 5 demonstrates substantially extended lead times distributed across much longer time periods, reflecting the impact of raw material procurement delays. In contrast, scenario 6 shows a marked reduction in batch lead times, with the distribution peak occurring around 100 days due to increased QA/QC capacity.

\newparagraph{Bottleneck analysis metrics}
Analysis of raw material stockouts (Appendix~\ref{s:appendix_additional_results} Table~\ref{tab:raw_material_stockouts}) reveals significant disparities between scenarios. In scenario 5, filters experience stockouts for 51.8\% of the days. Scenario 3 demonstrates moderate increases in enzyme stockouts, with average occurrences of 2.7 and 0.8 days per year respectively for both types of enzymes included, compared to baseline conditions with minimal stockout events.

\begin{figure}[h!]
    \centering
    \includegraphics[width=1\textwidth]{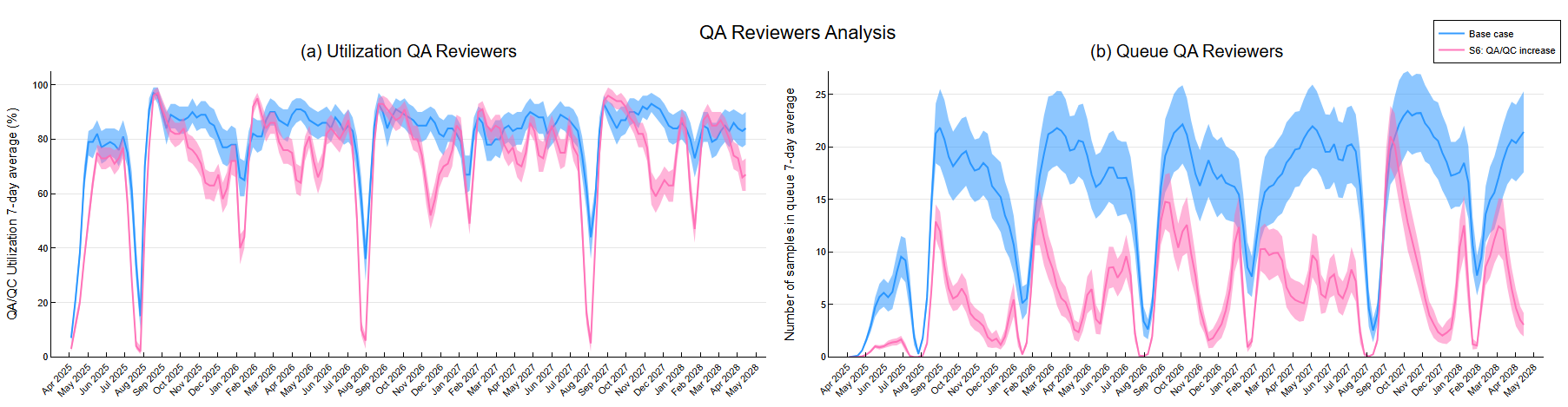}
    \caption{QA/QC personnel analysis comparing base case and scenario 6: (a) utilization rates, and (b) sample queue lengths.}
    \label{fig:combinedQaqc_bc_s6}
\end{figure}

Figure~\ref{fig:combinedQaqc_bc_s6} demonstrates that scenario 6 initially maintains utilization rates similar to baseline conditions, followed by progressive utilization reduction as the increased workforce processes accumulated sample backlogs faster. This shows that increased QA/QC capacity effectively reduces queues and thus increases total vaccine throughput.

\begin{figure}[h!]
    \centering
    \includegraphics[width=1\textwidth]{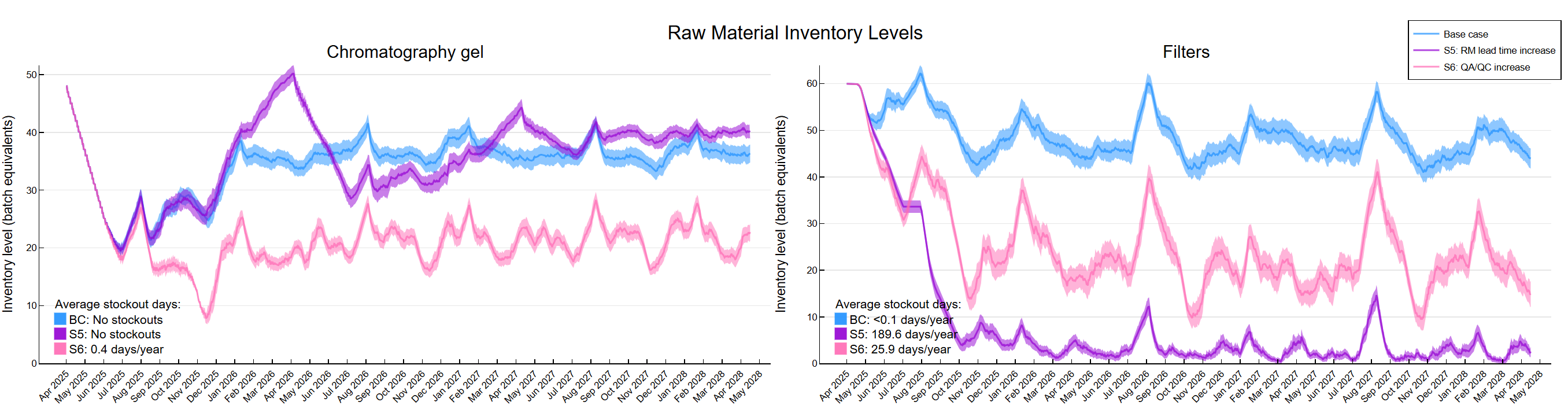}
    \caption{Inventory levels of two raw materials, chromatography gels and filters, of the base case, scenario 5 and 6.}
    \label{fig:rawMaterials_bc_s5_s6}
\end{figure}

When looking at the inventory levels of the raw materials of scenario 5 and 6 in Figure~\ref{fig:rawMaterials_bc_s5_s6}, we see for both raw materials a clearly different pattern compared to the base case. The inventory levels of chromatography gels drop quickly at the start and then fluctuate at a lower level (around 20 batch equivalents) for scenario 6, as more raw materials are used since the production rate is higher. For scenario 5, the inventory levels sharply increases, as the placed orders are arriving while the system is not producing due to stockouts of the filters.
The filters in scenario 6 show a similar seasonal pattern as the base case, but at a lower level (around 20 batch equivalents), but higher amplitude. This is due again to the higher production rate, which will lower the average stockpile, but also have a faster depletion rate and increased order quantities. For scenario 5, the inventory levels drop quickly at the start and then fluctuate near zero, indicating frequent stockouts. This is confirmed by the number of stockouts for scenario 5, as shown in Appendix~\ref{s:appendix_additional_results} Table~\ref{tab:raw_material_stockouts}, where filters are out of stock for 189 days per year (51.8\% of the days).

\subsection{Discussion}
\label{s:discussion}
Based on the results from the mRNA vaccine manufacturing simulation, several key insights emerge. 

The simulation reveals fundamental system dynamics that distinguish vaccine manufacturing from traditional production systems. Most notably, QA/QC personnel emerge as the primary bottleneck rather than production equipment, with utilization rates reaching 84.5\% on average under normal conditions while machine utilization remains below 33\%. This finding challenges conventional manufacturing assumptions where physical capacity typically constrains output, which was also identified by the work of \textcite{Lemmens2018}. The non-linear scaling effects observed in scenario 6 demonstrate that doubling QA/QC capacity yields substantial but diminishing returns (+79.1\% output improvement), as other system constraints begin to limit further scaling. Disruptions in raw material procurement (scenario 5) created stockouts during 51.8\% of production time, demonstrating how upstream constraints propagate through the entire manufacturing network. This behavior reflects the complex interdependencies characteristic of regulated pharmaceutical mRNA vaccine manufacturing, where multiple subsystems must operate in coordination to achieve system-wide performance improvements.

The analysis reveals distinct patterns of system resilience based on disruption characteristics, enabling a taxonomy of supply chain vulnerabilities. Transient disruptions, characterized by sudden onset but limited duration (scenarios 1 and 4), demonstrate the system's recovery capabilities through existing buffers and redundancies. These disruptions typically result in temporary performance degradation followed by a recovery within 6 to 9 weeks, as the underlying system structure remains intact.
Conversely, chronic disruptions that affect fundamental system parameters create sustained performance degradation that persists throughout the simulation period. The extended raw material lead times in scenario 5 created a permanent reduction in system capacity (-19.6\% after 36 months)
Lastly, workforce capacity reductions (scenario 2), a disruption occurring at the start of the simulation, created a prolonged recovery period (12 weeks). This highlights that early disruptions, when no buffers have been built up, can have disproportionate impacts on long-term performance. 

The simulation results offer evidence-based insights for policy development aimed at enhancing pandemic preparedness capabilities. 
The identification of QA/QC capacity as a strategic resource has important implications for regulatory policy and workforce development initiatives. Policymakers should consider targeted investments in QA/QC training programs and emergency protocols that enable rapid scaling of quality assurance capabilities during health emergencies. The simulation demonstrates that these interventions could yield substantial improvements in pandemic response capacity while maintaining safety standards.
Strategic stockpiling represents another critical policy consideration based on the raw material vulnerability patterns observed. Rather than focusing exclusively on finished product stockpiles with limited shelf life, the results support maintaining strategic reserves of stable raw materials that can be rapidly converted to vaccines for unknown future pathogens.

% ========================================================================================

\section{Conclusion}
\label{s:conclusion}

This study makes three significant contributions to vaccine manufacturing research and pandemic preparedness. First, we developed the first discrete-event simulation model that integrates production processes, QA/QC activities, and raw material procurement within pharmaceutical manufacturing, addressing a critical gap in existing studies that examine these components in isolation. Second, we created a decision-support system that enables policymakers and manufacturers to quantitatively evaluate disruption impacts and mitigation strategies under realistic operational constraints. Third, we conducted a detailed mRNA vaccine manufacturing platform case study that demonstrates the practical applicability of our integrated modeling approach and provides the first systematic dataset on mRNA vaccine manufacturing operations, compiled and validated with senior industry professionals, establishes benchmark parameters that can inform future vaccine manufacturing research and analysis.

The mRNA case study revealed critical insights into supply chain vulnerabilities and resilience patterns. QA/QC personnel emerged as the primary bottleneck, with utilization rates reaching 84.5\% compared to 33\% for production equipment, challenging conventional assumptions about manufacturing constraints. Raw material procurement disruptions proved particularly severe, with extended lead times reducing output by 19.3\% after three years of simulation and creating stockouts lasting 51.8\% of production time. The analysis demonstrated differential resilience patterns: acute disruptions (facility shutdowns, power outages) enabled rapid recovery within 6 to 9 weeks, while early or chronic disruptions (workforce shortages, persistent supply delays) created sustained performance degradation lasting from 3 months to the entire simulation period. Addressing our research question, the interaction between raw material constraints and QA/QC workforce limitations creates cascading and interacting effects that amplify system vulnerabilities, with the most effective mitigation strategy being targeted QA/QC capacity increase rather than production equipment expansion.

For manufacturers, our simulation model provides a systematic way for evaluating strategic investment priorities within their specific manufacturing contexts. The integrated modeling approach enables manufacturers to identify bottlenecks, assess supply chain configurations, and conduct scenario-based risk assessment before real disruptions occur. By modeling their own production processes, QA/QC activities, and raw material procurement systems, manufacturers can generate quantitative evidence to guide investment decisions tailored to their particular operational constraints and supply chain vulnerabilities. The model's flexibility allows adaptation to different vaccine platforms, and facility configurations, making it a versatile tool for strategic planning and operational optimization.

For policymakers, the framework supports evidence-based pandemic preparedness planning through quantitative assessment of regulatory interventions, strategic reserve requirements, and workforce development priorities. The model's ability to capture system-wide interdependencies makes it invaluable for designing coordinated response strategies that account for complex supply chain interactions rather than isolated interventions.

While this study presented a detailed discrete-event simulation model, several limitations should be considered. The independent scenario analysis approach does not capture compound disruption effects that may emerge when multiple disturbances occur simultaneously during severe pandemic conditions. The single-facility perspective, while enabling detailed operational analysis, does not address global supply network effects, technology transfer implications, or international coordination challenges that characterize real-world vaccine supply chains. Additionally, the model excludes economic considerations such as cost optimization, pricing dynamics, and financial constraints that significantly influence manufacturing decisions in practice.

Future research should address these limitations through several priority directions. Compound disruption analysis examining simultaneous failures (e.g., workforce reductions combined with supply delays) could provide insights into cascading failure patterns during severe pandemic conditions. Multi-facility network modeling could capture global supply chain dynamics, international coordination requirements, and distributed manufacturing strategies. Platform comparison studies examining mRNA, viral vector, and protein-based vaccine manufacturing could reveal technology-specific vulnerabilities and highlight possible opportunities. Furthermore, research could focus on deep analysis of raw material stockpile optimization strategies, incorporating stochastic demand patterns, and developing dynamic inventory policies that balance cost efficiency with supply security. Integration with economic modeling could enable cost-benefit analysis of resilience investments and support decision-making frameworks for both industry and policy applications. Machine learning and artificial intelligence applications could extend the model's utility by enabling automated identification of compound disruption risks, optimization of raw material procurement strategies, and early warning systems for emerging bottlenecks through pattern recognition in production and quality control data.

\paragraph{Acknowledgements}
This research has been funded by CEPI with reference VaxMan and a KU~Leuven C3-project with reference C3/25/003 - Het modelleren van het vaccin productie eco-systeem - VaxMod. We would like to thank Barbara Santini, Valerie Chambard and Matthew Downham from CEPI for their support and feedback throughout the project.

\paragraph{Declaration of generative AI and AI-assisted technologies in the manuscript preparation process.}
During the preparation of this manuscript, the authors used Claude Sonnet 4.5 to improve grammar and language quality. After using this tool, the authors reviewed and edited the content as needed and take full responsibility for the content of the published article.

\paragraph{Data availability}
Data will be made available on reasonable request to the corresponding author.

\paragraph{Conflict of Interests}
The authors declare no competing interests.

% ========================================================================================

\newpage

\printbibliography

\newpage
\appendix

\section*{Appendix}

\section{Complete case study diagram}
\label{s:appendix_complete_case_study_diagram}

\begin{landscape}
\begin{figure}[h!]
    \centering
    \includegraphics[width=1\linewidth]{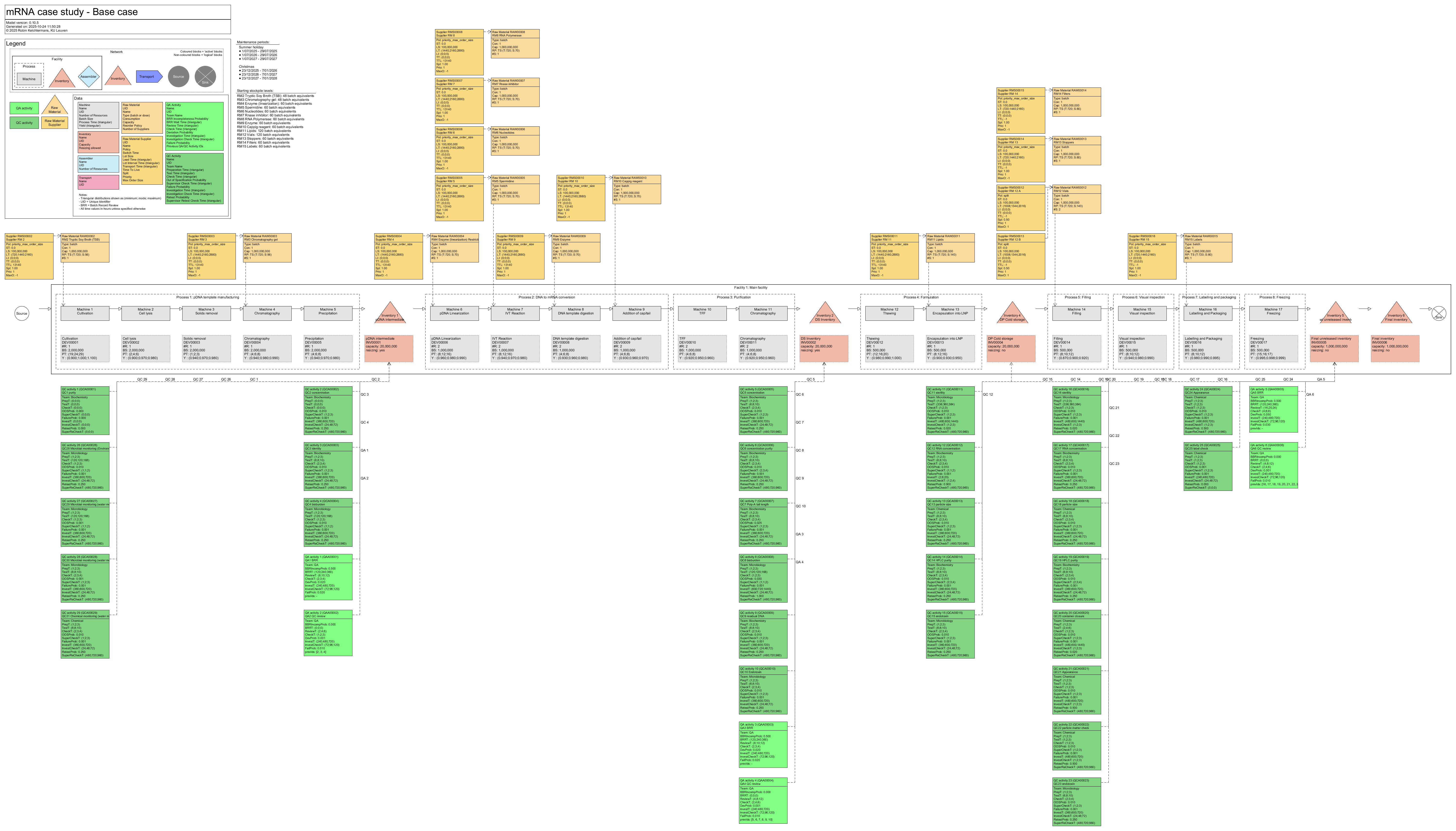}
    \caption{Complete case study diagram showing the different processes with their machines, inventories, QA/QC activities and raw materials.}
    \label{fig:complete_diagram}
\end{figure}
\end{landscape}

Figure~\ref{fig:complete_diagram} shows the complete case study diagram including all processes with their machines, inventories, QA/QC activities and raw materials. Each box contains all the data associated with that component.

\section{Additional results}
\label{s:appendix_additional_results}
Table~\ref{tab:raw_material_stockouts} shows the raw material stockout frequency (days per year) across all scenarios. In scenario 5, filters are out of stock for 189 days per year (51.8\% of the days), highlighting the severe impact of increased lead times on raw material availability. In scenario 6, many raw materials experience stockouts due to the significantly increased production rate, which depletes raw material inventories faster. This shows that the bottleneck shifts from QA/QC personnel to raw material availability when QA/QC capacity is increased.

\begin{table}[h!]
    \centering
    \caption{Raw material stockout frequency (days per year) across scenarios. No stockouts are indicated by a dash.}
    \begin{tabular}{lrrrrrrr}
        \toprule
        \textbf{Raw Material} & \textbf{BC} & \textbf{S1} & \textbf{S2} & \textbf{S3} & \textbf{S4} & \textbf{S5} & \textbf{S6} \\
        \midrule
        Tryptic Soy Broth & - & - & - & - & - & - & - \\
        Chromatography gel & - & - & - & - & - & - & 0.4 \\
        Enzyme (linearization) & 1.4 & 1.2 & 1.0 & 2.7 & 1.3 & 0.3 & 36.9 \\
        Spermidine & 0.2 & 0.5 & 0.4 & 0.4 & 1.1 & 0.2 & 47.3 \\
        Nucleotides & 0.8 & 0.7 & 0.7 & 0.6 & 1.1 & 0.4 & 48.0 \\
        Rnase inhibitor & 1.0 & 0.6 & 0.5 & 0.4 & 0.7 & 0.3 & 44.5 \\
        RNA Polymerase & 0.8 & 1.1 & 0.5 & 0.4 & 0.5 & 0.5 & 44.9 \\
        Enzyme & 0.5 & 0.5 & 0.7 & 0.6 & 0.6 & 0.4 & 23.6 \\
        Capping reagent & 0.2 & 0.3 & 0.3 & 0.4 & 0.3 & 0.2 & 27.8 \\
        Lipids & - & - & - & - & - & - & 0.7 \\
        Vials & - & - & - & - & - & - & - \\
        Stoppers & - & - & - & - & 0.1 & - & 26.0 \\
        Filters & - & - & - & - & - & 189.4 & 25.9 \\
        Labels & - & - & - & - & - & - & 25.2 \\
        \bottomrule
    \end{tabular}
    \label{tab:raw_material_stockouts}
\end{table}

\newpage

\end{document}